# Exploring the Potential of Two-Dimensional Materials for Innovations in Multifunctional Electrochromic Biochemical Sensors: A Review

Nadia Anwar[1, #], Guangya Jiang[2, #], Yi Wen[1], Muqarrab Ahmed[3], Haodong Zhong[1], Shen Ao[1], Zehui Li[4], Yunhan Ling[1], Grégory F. Schneider[2], Wangyang Fu[1, *], Zhengjun Zhang[1, *]

[1]School of Materials Science and Engineering, Tsinghua University, Beijing, 100084, P. R. China

[2]Leiden Institute of Chemistry, Faculty of Science, Leiden University, 2333CC Leiden, The Netherlands

[3]State Key Laboratory of Chemical Engineering, Department of Chemical Engineering, Tsinghua University, Beijing 100084, P. R. China

[4]School of Environmental Science and Engineering, Shanghai Jiao Tong University, Shanghai 200240, China

**Corresponding Author: Assoc. Prof. Wangyang Fu, Prof. Zhengjun Zhang,** School of Materials Science and Engineering, Tsinghua University, Beijing, 100084, P. R. China

**Email:** fwy2018@mail.tsinghua.edu.cn, zjzhang@tsinghua.edu.cn

**Abstract**

In this review, the current advancements in electrochromic sensors based on two-dimensional (2D) materials with rich chemical and physical properties are critically examined. By summarizing the current trends in and prospects for utilizing multifunctional electrochromic devices (ECDs) in environmental monitoring, food quality control, medical diagnosis, and life science-related investigations, we explore the potential of using 2D materials for rational design of ECDs with compelling electrical and optical properties for biochemical sensing applications.

## 1. Introduction

Electrochromism refers to a reversible change in the optical characteristics of a material induced by electrochemical reduction or oxidation. [1] This phenomenon was historically demonstrated by Berzelius and Wöhler through chemical reduction of yellow stoichiometric $WO_3$ to a blue material. [2] Electrochromism is typically realized using materials such as metal oxides, conjugated polymers, and organic compounds that can undergo reversible electrochemical reactions. These electrochromic (EC) materials exhibit distinctive visible color changes from bleached to opaque and can maintain their coloration even after the electrochemical voltage source is disconnected, leading to their use in innovative applications such as supercapacitors, smart windows, and sensors. [3-5]

In the production of EC biochemical sensors, an EC material layer is deposited on an electrode with a biochemical recognition element immobilized on top. This configuration allows multimode registration of analytical signals, which is applicable to a broad spectrum of biological and chemical analyses. [6] Such sensors are easily activated by simple voltage sources and provide direct optical readout of the sensor response, thereby saving energy and simplifying the electrochemical instrumentation. Ideal EC materials for these applications should possess rapid response times, good stability, and excellent visual contrast between their colored states. [7] Traditional materials such as conjugated polymers, metal oxides, and viologens are commonly utilized due to their accessibility, ease of fabrication, and customizable properties. [8]

Two-dimensional (2D) materials such as graphene and molybdenum disulfide ($MoS_2$), which have unique electronic and optical properties and increased surface accessibility for analytes due to their reduced dimensionality, have also been explored. [9] Despite the challenges in synthesizing high-quality, large-area, homogeneous 2D materials and integrating them into devices, the potential of 2D materials has spurred significant research efforts in investigating their physical and chemical properties for next-generation biochemical sensing applications, which require high sensitivity and quick response times. [10-15]

This review critically summarizes the progress in EC sensors based on 2D materials. We discuss the surface functionalization of 2D materials and how it affects their electrical band structures, optical properties, electrochemical performance, and overall sensing capabilities. The current trends in and prospects for deploying multifunctional EC biochemical sensors in environmental monitoring, food quality control, medical diagnosis, and life sciences are also evaluated. As demonstrated in Figure 1, the optimal performance of EC devices (ECDs) for biochemical sensing is achieved through two key strategies: tuning or modification of the surface properties of the 2D material to enhance its interaction with target molecules and rational design of ECDs to achieve the desired electrochemical and optical properties. By integrating the advantages of 2D materials with sophisticated device design, our goal is to develop EC biochemical sensors that have both a high sensitivity and a quick response.

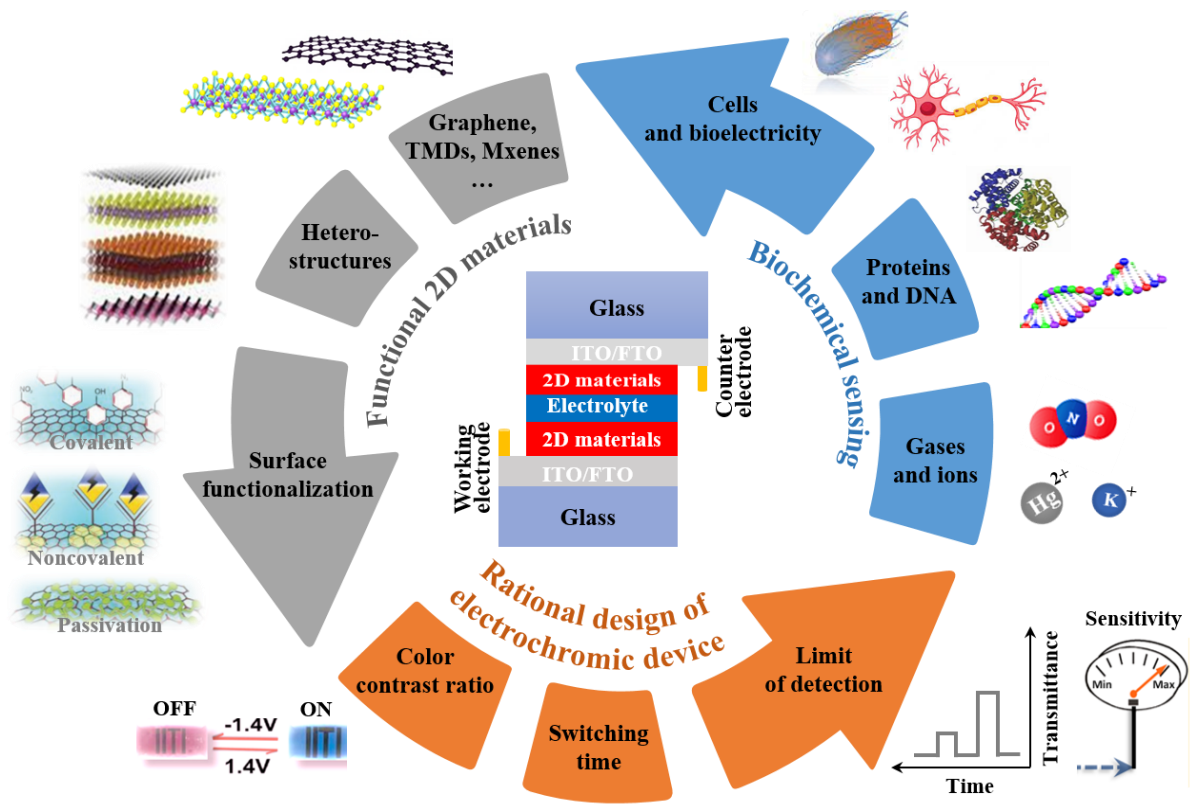

**Figure 1:** Summary of EC biochemical sensors based on 2D materials.

## 2. Structure and working mechanism of ECDs based on 2D materials

The utilization of EC materials in biochemical sensors capitalizes on the reversible alteration in optical properties that occurs during electrochemical reduction or oxidation to detect biological or chemical stimuli. Since the introduction of EC technology by J.R. Platt in 1961 and the development of the first ECD by Deb et al. in 1969, significant advancements have been made in the field. [16] These improvements have been driven by the integration of various detection techniques, such as chemiluminescence, photoluminescence, electrochemiluminescence, chromatography, mass spectrometry, surface-enhanced Raman scattering, and colorimetry. [17] Recently, ECDs employing 2D materials have attracted considerable attention. These materials are chosen for their distinctive electrical and optical properties, which are highly advantageous for developing advanced EC sensors. [18, 19]

**2.1. Synthesis of emerging 2D EC materials**

Traditional EC materials primarily consist of inorganic substances, such as $WO_3$ and other transition metal oxides ($MoO_3$, $TiO_2$, $V_2O_5$, etc.), and organic materials, including conducting polymers (polypyrrole, polythiophene, polyaniline, and their derivatives), metal complexes, and small organic compounds such as the viologen family, metal phthalocyanine, and Prussian blue derivatives. The range of EC materials has recently been expanded to versatile 2D materials, significantly revitalizing research in this field. Since 2014, reports on the EC effects of various 2D materials, such as reduced graphene oxide (rGO), coordination nanosheets (CONASHs), copper chalcogenides, covalent organic frameworks (COFs), transition metal carbides/nitrides/carbonitrides (MXenes), and hydrogen-bonded organic frameworks (HOFs), have sparked renewed interest. 2D materials differ from traditional EC materials in their unique physical and chemical properties, which can markedly improve the EC performance. For example, their extremely high specific surface area and broad van der Waals gap facilitate faster ion transport at both the micro- and macroscales, enhancing the switching speed of ECDs to meet diverse requirements. This section summarizes recent developments in novel 2D EC materials, with an emphasis on understanding their structure-property relationships.

Emerging 2D EC materials can be fabricated using either top-down or bottom-up methods. [20] In the top-down approach, a bulk material is progressively reduced in size or modified to achieve the desired properties or structures. Common techniques utilized in this method include machining, milling, or lithography, which involve removing or shaping a material on a larger scale. This approach is well suited for creating macroscopic or mesoscopic structures and devices, although it presents challenges in achieving custom functional designs. In contrast, in the bottom-up approach, materials are built from atomic or molecular components, which are gradually assembled into larger structures. Typical techniques employed in this method include self-assembly, chemical synthesis, or deposition, which enable the creation of materials at the atomic or molecular scale. These smaller components then aggregate, forming more extensive structures. The bottom-up approach offers precise control over the material properties at the atomic or molecular level, facilitating the production of nanoscale and tailored materials. In practice, these approaches often need to be combined to achieve the specific material properties and structures desired for 2D EC materials.

The bottom-up approach in nanoparticle synthesis begins with the use of precursor materials, and chemical reactions are carefully controlled to yield nanoparticles with specific sizes, shapes, and compositions. This method ensures precise control over the physical and chemical characteristics of the final product. Conversely, the top-down approach in microchip fabrication involves the creation of intricate circuit patterns on a silicon wafer. This is achieved by etching unwanted material using a series of photolithography and etching steps. For 2D materials that have bulk analogs, mechanical forces, sonication, chemical or electrochemical reactions, and interlayer contacts or chemical bonds to realize intercalation are utilized in top-down fabrication, although adjustment of the properties through structural alteration poses challenges. In contrast, in the bottom-up approach, various synthetic techniques are leveraged to create 2D materials from fundamental building blocks such as metal atoms and ions, ligands, and molecules. Techniques such as liquid–liquid interfacial synthesis, chemical vapor deposition (CVD), sol-gel methods, electrodeposition, electrospinning, and hydrothermal methods are employed. This approach not only facilitates the synthesis of materials at the nanoscale but also offers substantial flexibility in modifying the structures and properties of synthetic materials through strategic design and component selection, as illustrated in Figure 2a.

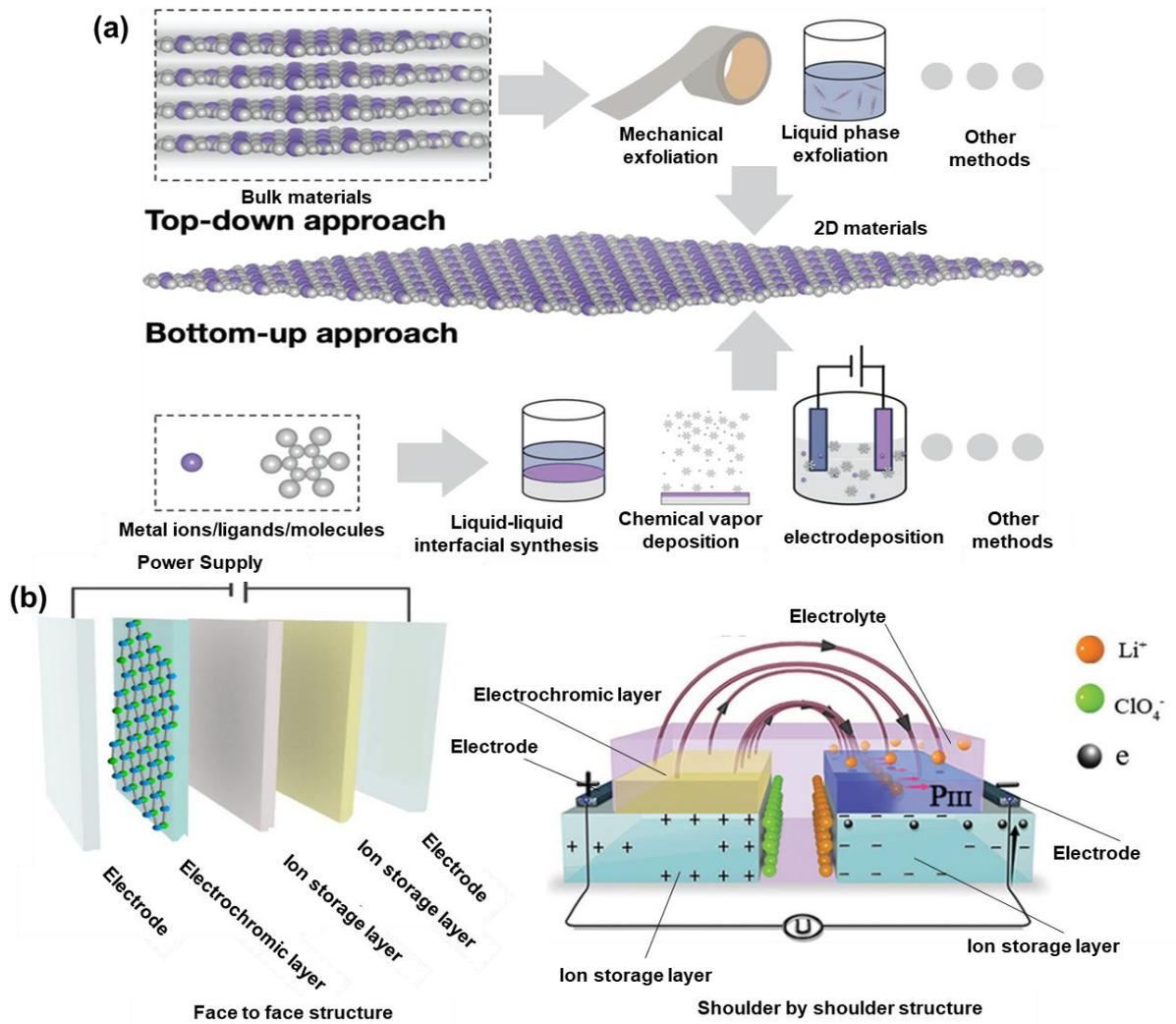

**Figure 2**: Design of and fabrication methods for ECDs. (a) Synthesis of emerging 2D EC materials via top-down and bottom-up approaches. [21] (b) Structures of traditional layered and lateral ECDs with five functional levels, including two electrodes, an ECL, an ion transport layer (ITL), and an ion storage layer (ISL). [22] Evolution of ECDs from 'face-to-face' to 'shoulder-by-shoulder'. [23]

## 2.2. ECD architecture

As illustrated in Figure 2b (left panel), the fundamental architecture of an ECD begins with its substrate. This substrate serves as the base material on which the device is constructed. It is typically transparent to ensure visual functionality and provides the necessary mechanical support. Common substrates used in ECDs include glass or transparent conductive materials such as indium tin oxide (ITO).

Electrode layers, which are crucial for the functioning of the ECD, are deposited on top of the substrate. These layers enable the application of an electric field across the device. ECDs generally incorporate two types of electrodes: a transparent conducting electrode and a counter electrode. The transparent conducting electrode, which is often made of a material such as ITO, is designed to be transparent and is usually positioned closer to the viewer to allow light to pass through the device.

The EC layer (ECL), positioned between the electrodes, is the active component of the device. This layer undergoes a reversible color change in response to the electric field applied across the electrodes. It is typically fabricated from a thin film of an EC material, such as $WO_3$ or viologens. Upon the application of a voltage across the electrode layers, the optical properties of the ECL are adjusted, thereby changing its color or level of transparency.

The ion-conducting layer is a critical component of ECDs and is positioned between the ECL and the counter electrode. This layer facilitates the movement of ions, enabling them to travel between the ECL and the counter electrode upon the application of an electric field. Reversible ion movement is essential for altering the color of the ECL. Ion-conducting materials commonly used in this layer include polymer electrolytes or gel electrolytes, which provide good ionic conductivity and mechanical flexibility.

The counter electrode opposite to the transparent conducting electrode serves as a vital component by completing the electrical circuit within the ECD. Unlike the ECL, the counter electrode is typically made from materials that do not change color. Common choices for counter electrode materials include platinum or conductive polymers, which offer durability and stable electrochemical properties.

In some ECD designs, an additional electrolyte layer is included to further facilitate the efficient movement of ions between the ECL and counter electrode. Depending on the specific requirements and design of the device, this electrolyte can be either a liquid or a solid-state material, each of which offer different benefits in terms of device performance and integration.

To protect the ECD from environmental factors such as moisture and oxygen, which can degrade materials and impair device functionality, an encapsulation layer is often applied. This protective layer is typically composed of a thin film of a barrier material such as silicon dioxide ($SiO_2$) or aluminum oxide ($Al_2O_3$). These materials are chosen for their excellent barrier properties, which help ensure the longevity and reliability of the ECDs.

Importantly, the configuration of an ECD can significantly vary based on its specific application and design requirements. The choice of materials, thickness of layers, and deposition techniques employed can be tailored to meet specific objectives, such as the desired color range, response time, and overall device performance. For instance, the use of miniaturized structures with porous electrodes can markedly enhance molecular diffusion and interactions, thereby optimizing the performance without detrimental effects. One notable example is the use of 2D COFs. These materials are characterized by a porous structure and interlayer π-π interactions, which contribute to their excellent optoelectronic properties. The adjustable porosity and specifically designed molecular structure of COFs offer extensive possibilities in the field of optoelectronic materials and biosensors.

Sandwich configurations are typically adopted for ECDs because they minimize internal resistances and reduce the switching time, thereby enhancing the performance, efficiency, and optical properties. Furthermore, this structure supports the integration of multiple functional layers within a single device, enhancing both its functionality and versatility. [24] However, there are inherent drawbacks to this compact structure, particularly when transparent but impermeable electrodes are employed. This configuration can slow diffusion processes and may impede the interaction between the analytes of interest and 2D materials, thus conflicting with the objective of achieving highly sensitive and rapid detection. To address these issues, an alternative lateral layout with open sensing windows, as depicted in Figure 2b (right panel), could be more advantageous. This layout typically includes several key components that collaboratively function to achieve the desired EC effect. In such lateral configurations resembling a planar layout, a novel approach to EC operations can be realized. For example, when conductive ITO layers are used, they can be functionalized as two split capacitance plates when both films are transparent. Upon the application of a working voltage, Li ions are propelled by the edge effect of the electric field and intercalate into the layers at a negative potential. This process results in an uneven color distribution, with more Li ions being injected into the center of the two EC films than into other areas due to the significant edge effect of the electric field.

**2.3. ECD working mechanism**

The functionality of an ECD is primarily determined by the specific materials used and its design. Commonly, ECDs such as EC windows can control the amount of light entering a room through a change in their opacity. These devices generally consist of two conductive glass layers separated by an electrolyte and a layer of an EC material, such as $WO_3$. When a voltage is applied, ions within the electrolyte move through the ECL, triggering a chemical reaction that alters the color and transparency of the device.

EC displays, which are another type of ECD, are utilized in electronic devices such as smartphones or e-readers to display text or images. These displays typically employ an EC polymer or a small molecule as the active material, which changes color upon voltage application. The inclusion of dopants facilitates charge transfer during bias-induced redox switching, enhancing the switching speed. As shown in Figure 3a, the charge dynamics and redox reactions between the electrodes play a crucial role in the color change of the current ECDs, as in devices using ethyl viologen (EV) and polythiophene (P3HT) materials. [25] When a negative voltage (-1.4 V) is applied to the $MoS_2$ electrode (Fig. 3a), $EV^{2+}$ starts to absorb electrons from the electrode, resulting in a blue color, while P3HT is simultaneously oxidized from its neutral magenta state to a transparent polaronic state. These phenomena result in the device appearing blue until the voltage is discontinued.

The mechanism of these devices involves the movement of electrons between the active material and an electrode, which causes a change in the absorption or reflection of light (Figure 3b). In electrochemical reactions, mediating electrons can be exchanged to facilitate electron transfer from one molecule to another. The electron transfer process can directly occur between reactants or be facilitated by a mediator that helps transfer electrons between reactants. [26] Overall, the ECD mechanism involves the controlled movement of ions or electrons between different layers of the device, resulting in a change in optical properties.

This review article also focuses on scaling and on solid electrolytes to achieve a significant sensing response for the future development of EC biochemical sensors. Taking an ECD based on 2D materials as an example, [27] Figure 3c and Figure 3d present optical images of the ECD biased at 0 V and 5 V, respectively. The device in the images has a university logo behind it, and the logo is visible at 5 V. The mechanism of this EC behavior is further detailed in the schematic diagrams in Figures 3e–3g. When a voltage below 2.5 V is applied, this voltage primarily polarizes the ionic liquid, leading to an "accumulation" phase in which charges build up at the interfaces between graphene and the electrolyte. This interaction results in minimal optical modulation, approximately 2% of the initial transmittance, due to doping of the graphene layers at these interfaces. However, application of a voltage greater than 2.5 V introduces structural flaws and initiates the ion intercalation process within the graphene layers.

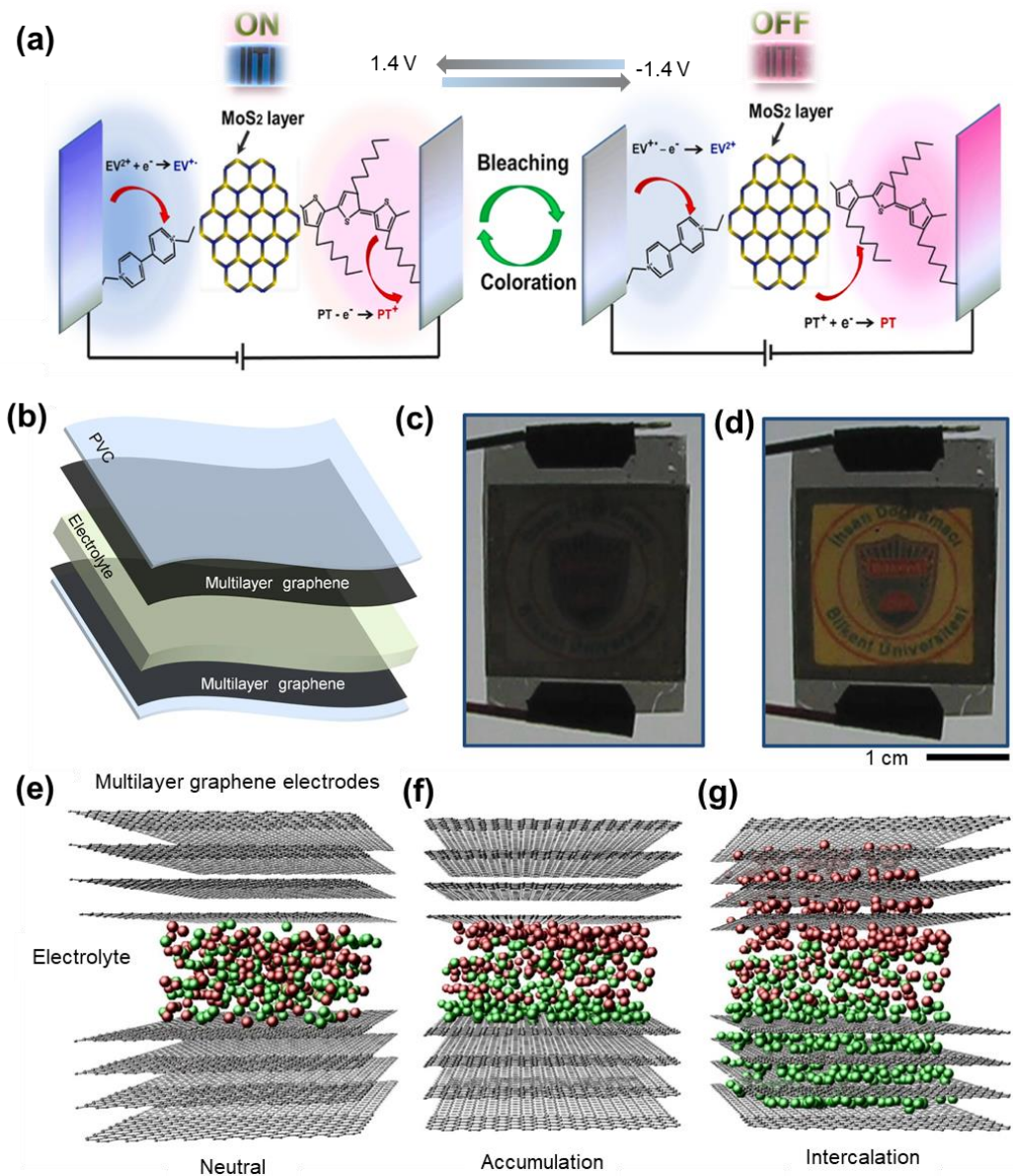

**Figure 3:** Working mechanism of a device when applying a voltage. (a) Schematic illustration of the mechanism illustrating the mobility of electrons during the coloring (magenta to blue) and bleaching (blue to magenta) processes. [25] (b) Illustration of a graphene ECD from an exploded perspective. The device is created by fusing two polyvinyl chloride (PVC) substrates with graphene coatings and injecting an ionic liquid into the space between the graphene electrodes. (c and d) Photos of the device taken at bias voltages of 0 V and 5 V, respectively. The Bilkent University logo can be seen at 5 V. The demonstrated device consists of two multilayer graphene electrodes with 78 layers each produced at 925 °C. (e-g) Schematic illustration of the three suggested operating regimes: intercalation, charge accumulation, and neutral. [27]

The transmittance spectra of the ECD exhibit significant variation under applied voltages ranging from 0 V to 5 V, as shown in Figure 4a. [27] At 0 V, the transmittance is minimal, approximately 8%, and slightly varies with wavelength. The transmittance dramatically increases to 55% at 900 nm when the voltage is increased to 5 V, demonstrating the profound impact of voltage on the optical properties of the device (Figure 4b). The spectrum is primarily modulated due to the electrostatic doping that blocks the interband transitions in the graphene layers, which necessitates efficient doping of the underlying layers for optimal functionality. A high bias voltage enables doping of the upper layers, with ion intercalation into the graphene layers facilitating effective doping. The operation of the device, in terms of transmittance modulation and charging time, appears stable even when using a thicker spacer containing 100 µL of an ionic liquid. However, the use of smaller spacers with a lower ionic liquid volume in large-scale devices can create electrical imbalances between the top and bottom graphene layers, potentially affecting the device performance. The typical hysteresis behavior observed in supercapacitors due to the development of electric double layers is also noted in the charging and discharging cycles (Figure 4c). [28]

Tang et al. [23] constructed a $WO_3$ *vs.* $WO_3$ lateral (shoulder-by-shoulder) structure EC system and systematically discussed its EC properties and superior application potential. For the $WO_3$ *vs.* $WO_3$ lateral structure EC system, the coloring and bleaching of the EC units (from 40% to 85%) require approximately 350 s and 180 s, respectively, at a driving voltage of 1.2 V, which is completely impossible to achieve in the traditional vertical (face-to-face) structure (Figure 4d). At the same time, the $WO_3$ *vs.* $WO_3$ shoulder-by-shoulder ECD also has excellent durability, and there is no deterioration or bubbles after 15000 processes at a driving voltage of 1.2 V (Figure 4e). Moreover, the shoulder-by-shoulder structure provides greater flexibility in material selection, shape, and placement of the counter electrode, enhancing the device applicability across various fields.

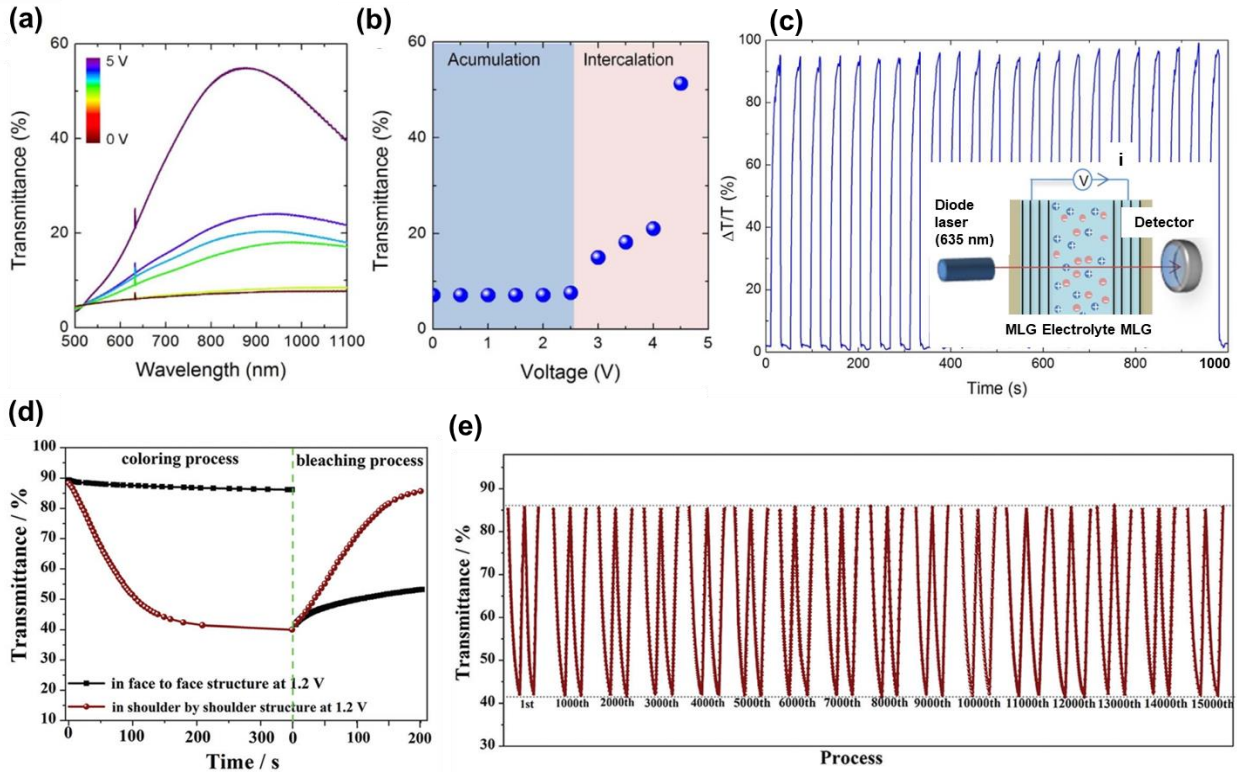

**Figure 4:** Performance of face-to-face and shoulder-by-shoulder ECDs. (a) Device transmittance spectra at 0 to 5 V. (b) Transmittance variation at 800 nm with the applied voltage. (c) Experimental configuration utilized to characterize graphene ECDs both electrically and optically. Features of the device optical switching at 635 nm. [27] (d) Comparison between the traditional face-to-face and shoulder-by-shoulder structures at a low working voltage of 1.2 V. (e) Cyclic testing to determine the durability of the device at 550 nm (15000 processes) at +1.2 V for 360 s and -1.2 V for 200 s. Note that after each set of 1000 processes, the cell was switched off for the transmittance test. [23]

## 2.4. Electrolytes in ECDs

The choice of electrolyte material for an all-solid-state ECD is determined by the specific requirements of the device and the materials used in its construction. Common types of solid-state electrolytes in ECDs include solid polymers, ceramics, glass ceramics, and inorganic salts. The choice of the electrolyte material can influence factors such as the conductivity, ionic mobility, stability, and compatibility with the other components of the device. [29]

Gel polymer electrolytes (GPEs) are particularly favored in all-solid-state ECDs due to their unique properties. They can form a solid-like structure that still permits ionic conduction, making them ideal for applications that require both stability and efficient ion transport. GPEs are usually positioned between the electrodes and ion-storage layer, effectively forming the core functional structure of the ECD.

Ceramic solid electrolytes, in comparison, exhibit higher ionic conductivity and mechanical stability than GPEs. However, their rigidity requires different fabrication methods, such as solid-state sintering or screen printing. These methods can create denser and more homogeneous layers of the ceramic electrolyte, which can significantly enhance the overall performance of the ECD. Despite their advantages, the lack of flexibility of ceramic electrolytes may limit their use in devices that require bending or flexibility. [30]

**3. Performance of ECDs based on 2D materials**

2D materials such as graphene and transition metal dichalcogenides (TMDs) continue to attract significant attention in the ECD field. These materials are renowned for their unique electronic and optical properties that allow the color or transparency of ECDs to be altered in response to applied voltages or currents. The performance of these devices is influenced by several factors, including the type and quality of the 2D material used, device architecture, and operating conditions. The key performance metrics for these devices are the optical contrast—i.e., the difference in the optical density between colored and uncolored states—and switching speed—i.e., the time it takes for the device to switch between states.

Recent studies suggest that ECDs based on 2D materials can achieve high optical contrast and fast switching speeds. For example, graphene-based devices have been reported to demonstrate optical contrasts of up to 80% with switching times of less than 1 s. TMD-based devices have also shown promising results, achieving optical contrasts of up to 90% and switching times under 10 s. Additionally, the stability, durability, and scalability of these devices are crucial for their practical application, making 2D material-based ECD performance a vibrant area of ongoing research.

Before delving into the multifaceted performance parameters, examining a representative case is beneficial. $MoS_2$, which is a typical anodic EC material, illustrates the intricate EC mechanisms and performance enhancements possible with 2D materials. [31] When Li+ intercalates into the MoS2 lattice, it forms $LiMoS_2$, leading to charge transfer predominantly from $Mo^{+6}$ to $Mo^{+4}/Mo^{+5}$. This hypothesized mechanism suggests the complex interplay within the material structure that affects its EC performance. Factors such as the coating thickness, annealing effect, and choice of solvent critically impact the performance of $MoS_2$ in ECD applications. Significantly, when $MoS_2$ is used in conjunction with $WO_3$ as a counter electrode, the performance of $MoS_2$-based ECDs markedly improves. Under the cyclic application of a potential between 0 V (bleaching) and 1.5 V (coloring), $MoS_2/WO_3$ ECDs exhibit a stable performance for up to five cycles without notable degradation. At these low potentials, MoS$_2$/WO$_3$ ECDs exhibit an optimal optical contrast, reaching approximately 75%. While pure MoS$_2$-based ECDs show limited stability over multiple cycles, the incorporation of WO$_3$ enhances both their performance and stability. The switching times recorded for MoS$_2$/WO$_3$ ECDs are 5.1 s for coloring and 4.3 s for bleaching at low voltages. In terms of coloration, pure MoS$_2$ ECDs display a dark green color due to the light absorption characteristics of the ferrocene electrolyte at a wavelength of 680 nm (red spectrum). In contrast, the MoS$_2$/WO$_3$ combination produces a dark bluish-green color, indicating significant modification of the visual output.

## 3.1. Contrast ratio and optical modulation

The effectiveness of dyes in ECDs is predominantly evaluated based on the optical modulation [32] and contrast ratio. The contrast ratio refers to the difference in the absorbance or transmittance at a specific wavelength between the colored and bleached states of the device. High optical modulation and contrast ratio are crucial for enhancing the visibility and effectiveness in applications such as biomedical devices.

Optical transmittance measurements from 300 to 900 nm were conducted to compare the performance at 0 V (bleached state) and -1.5 V (colored state). As depicted in Figure 5a, all the films demonstrate low transmittance at -1.5 V and high transmittance at 0 V, illustrating significant optical modulation. Notably, the MoS$_2$-WO$_3$ (MSW) nanocomposite ECD films exhibit an optical modulation ranging between 75% and 60% at 700 nm. This value is substantially higher than that of the standalone WO$_3$ films, which show approximately 53% modulation, underscoring the benefit of integrating MoS$_2$ to enhance the optical properties.

The behavior of these devices under various electrical potentials was also investigated. As shown in Figure 5b, the color changes induced by applying a potential are reversible. In their normally bleached state, the devices exhibit a yellow color due to the typical ferrocene absorption. However, upon the application of a 1.5 V potential, this yellow color changes to a dark blue color in the colored state, which reverts to yellow when the potential is reset to 0 V. Among the various compositions, the MSW2.0 ECD displays the most pronounced blue color at 1.5 V, followed by the MSW0.5 and MSW0.2 ECDs.

Figure 5c shows the changes in the EC transmittance at 700 nm as a function of the applied voltage, revealing a linear decrease in the transmittance with decreasing voltage. Understanding the kinetics of coloration and bleaching is essential for practical applications. The switching times for the WO$_3$ film are 12.5 s for coloration and 9.0 s for bleaching. In contrast, the MSW ECDs show slightly longer switching times, ranging from 10.0 to 11.0 s for coloration and from 8.0 to 8.8 s for bleaching. These times are slightly longer than those of other previously reported WO$_3$-based ECDs, for which 90% coloration occurred in only 2.5 s and 90% bleaching occurred in 7.6 s. [31]

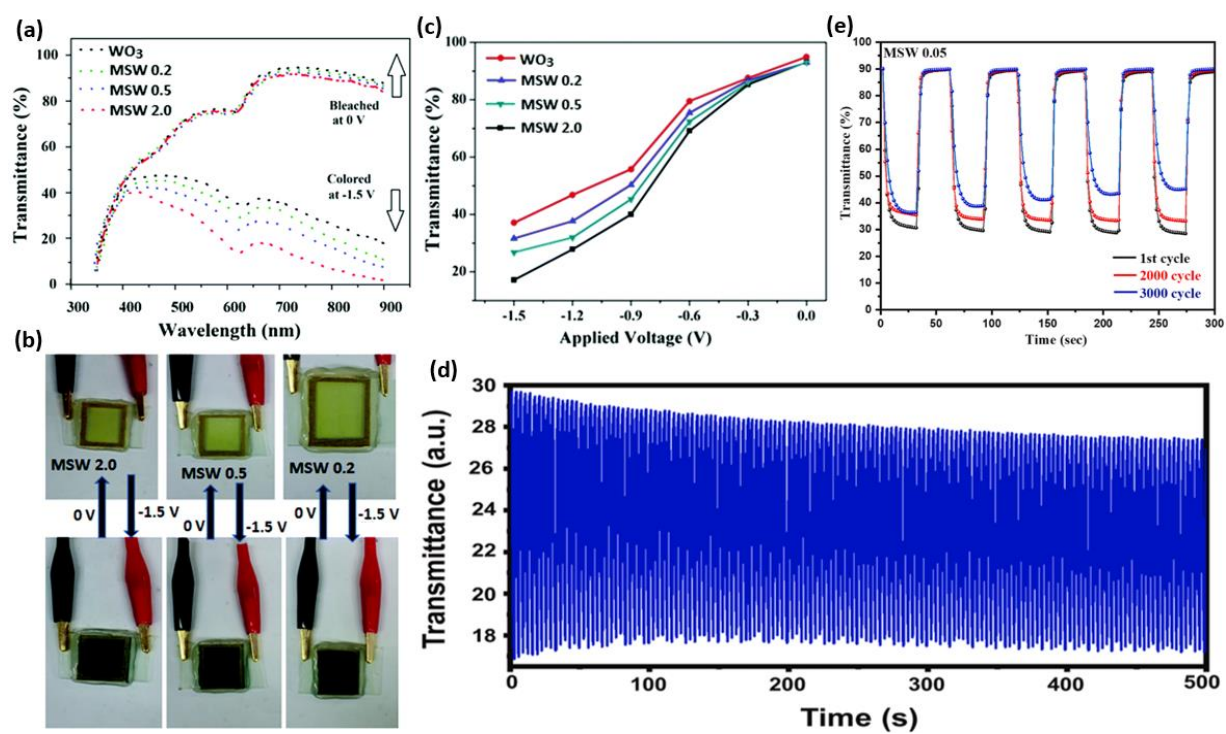

**Figure 5:** Performance of 2D material-based ECDs: (a) WO$_3$ and MSW film optical transmittance *vs*. wavelength curves at 700 nm with a potential of 0 V (bleached state) and 1.5 V (colored state) applied. (b) Digital images of the colored and bleached MSW film states. (c) Variation in transmittance at 700 nm for the WO$_3$ and MSW films with applied voltage. [31] (d) Device optical switching kinetics when repeatedly switched between 1.4 V and -1.4 V. [25] (e) Stability of the MSW0.05-based flexible ECDs (1st, 2000th, and 3000th cycles). [33] The performance of the ECD is given in Table 1.

**Table 1:** Performance of ECDs.

| Device Composition | Color Change | Optical Contrast (%) | Coloration Efficiency (cm$^2$ C$^{-1}$) | Switching Time/s Coloration $t_c$ (s) | Switching Time/s Bleaching $t_b$ (s) | Cycling Stability | Ref. |
|---|---|---|---|---|---|---|---|
| **P3HT/MoS$_2$+viologen** | magenta to blue | 42 | 580 | 0.8 | 0.4 | 100 | [25] |
| **PV–rGO/FTO** | purple to colorless | 43 | 142 | 6 | 9 |  | [34] |
| **MoO$_3$/ITO** | dark blue to transparent | 48.6 | 161.4 | 9.8 | 11.3 | 1000 | [35] |
| **MoSe$_2$/WO$_3$** | gray to white | 74.61 | 62 | 3.5 | 3.7 | 3000 | [33] |
| **MoS$_2$ QD/WO$_3$** | dark bluish to light green | 75 | 56.2 | 5.1 | 4.3 | 1000 | [36] |
| **rGO/NiO** | black to transparent | 53 | 30.5 | 3.4 | 5.3 | 1000 | [37] |
| **MoS$_2$/WO$_3$** | yellow to dark blue | 75 | 51.2 | 11.0 | 8.8 | | [31] |
| **MoS$_2$/WO$_3$** | yellow to dark blue | 84 | 67 | 19 | 9 | 100 | [4] |

### 3.2. Coloration efficiency

The coloration efficiency (CE) is a critical metric of ECDs and represents the relationship between the optical density (OD) and optical absorbance (A) at a specific wavelength per unit area of charge. [38] This metric is crucial for evaluating the efficiency of electronic color-switching processes, particularly in terms of power consumption and regulatory compliance. The CE essentially quantifies how effectively the coloration of an ECD can be modulated in response to electrical input.

Due to their unique characteristics, such as an atomically thin structure and strong van der Waals bonds, 2D materials often exhibit high CEs. These properties allow rapid and efficient charge transfer and color change, making these materials ideal for use in high-performance ECDs. For example, the CE of fabricated ECDs was extracted from the slope of the curves, and the CE of the MSW0.05-based ECD decreased from 59 cm$^2$ C$^{-1}$ to 45.7 cm$^2$ C$^{-1}$ after 10,000 cycles. [33]

### 3.3. Switching time

The switching time is a pivotal characteristic for assessing the performance of ECDs. It essentially measures the speed of the devices in response to electrical stimuli and is divided into two components: coloring time ($t_c$) and bleaching time ($t_b$). These metrics are crucial for applications in which rapid changes are desirable, such as in smart windows, displays, and various signaling devices.

A notable example of rapid switching can be observed in a device that switches from magenta to blue. [25] The device achieves 77% of its maximum absorbance in only 0.8 s during the coloring phase. This speed is sufficient for the color change to be distinctly noticeable. Conversely, during the bleaching phase, in which the device transitions from blue back to magenta, it reaches its peak absorbance of 67% in only 0.4 s. This exceptionally fast response time, less than half a second for both transitions, positions this device as one of the quickest among all-organic solid-state ECDs. The factors influencing the switching time include the ionic conductivity of the electrolyte and the chemical composition, crystal structure, morphology, and microstructure of the active materials. [30]

## 3.4. Long-term stability and durability

The long-term stability and durability are essential characteristics for evaluating the practical usability of ECDs, especially for applications that require frequent switching or prolonged usage. To accurately assess the endurance and reliability of these devices, operational tests involving continuous and repetitive cycling are conducted.

In a representative device stability test, [25] the device was subjected to a symmetric rectangular pulse voltage of 1.4 V, with each polarity applied for 3 s for a total cycle duration of 6 s. This test protocol was specifically designed to simulate real-world conditions in which the device would need to repeatedly switch between states. The cyclic absorbance at a wavelength of 515 nm was meticulously recorded, as illustrated in Figure 5d. The data from this test reveal that the device exhibits commendable cyclability, maintaining a consistent performance over more than 100 cycles. Notably, there is only a 3% decrease in the maximum transmission after 500 s of continuous switching. Furthermore, the performance remains relatively stable for the duration of the test period.

With a high durability, the ECD can withstanding frequent changes without significant degradation, indicating that it is suitable for various applications, such as in smart windows, adaptive eyewear, and variable-opacity displays.

## 3.5. Mechanical stability

The advancement of flexible electronics has set new benchmarks for the performance criteria of ECDs, particularly in terms of mechanical stability. ECDs designed for flexible applications must maintain stable optical properties even when subjected to mechanical deformations such as bending, stretching, or twisting. This demand underscores the need for devices that can withstand physical stresses without compromising their functionality.

In sandwich-structured ECDs, different layers experience various degrees of stress during mechanical deformation. Typically, the outer layers of the devices are subjected to tensile stress, while the inner layers may experience compressive stress. [39] Furthermore, a single layer can simultaneously undergo both compressive and tensile stresses, particularly near deformation points. This complex stress distribution can significantly affect the device performance by altering the conductivity of the transparent conductors and affecting ion diffusion in the electrolytes and electrodes.

An illustrative example can be seen in the performance of MSW0.05-based flexible ECDs. [33] These devices demonstrate robust transmittance and good optical qualities under mechanical stress, as shown in Figure 5e. Transmittance curves captured at a wavelength of 700 nm, with the applied voltage ranging from 1.2 V (colored state) to 0 V (bleached state), showcase their effective functionality. Additionally, these ECDs display commendable stability over 3000 cycles, indicating their suitability for applications requiring durable and flexible EC solutions.

## 4. Multifunctional applications of ECDs

The optical properties of an ECD, such as the color or transmittance, can change when an electrical voltage is applied. This change is due to the materials undergoing reversible electrochemical reactions in an electric field. ECDs are versatile; they are used in smart windows, displays, mirrors, and sensors and can be easily controlled by simple electrical signals. They are also used in multifunctional applications; for instance, when combined with solar cells or energy storage systems, they can be used to create smart windows that not only exhibit optical property modification but also generate or store energy. Additionally, EC materials can be incorporated into sensors that respond to environmental changes such as temperature, humidity, or the presence of specific gases or chemicals. Due to their broad applicability and ease of integration, ECDs and their materials are highly valuable across various technologies.

### 4.1 Traditional applications of ECDs

EC supercapacitors integrate energy storage and color-changing capabilities in a single device. These devices exhibit typical pseudocapacitive behavior, as evidenced by nonlinear galvanostatic charge–discharge (GCD) curves of NiO and rGO/NiO films across various current densities (0.5–2.5 mA $cm^{-2}$). At a current density of 1 mA $cm^{-2}$, the NiO films show a columbic efficiency of 77%, while the rGO/NiO films achieve an efficiency of 86.7%, as shown in Figure 6a and Figure 6b. The maximum capacitances of 184 mF $cm^{-2}$ for NiO and 269 mF $cm^{-2}$ for rGO/NiO are recorded at 0.5 mA $cm^{-2}$, with the reductions in capacitance at higher current densities likely due to poor electrolyte diffusion and penetration (Figure 6c). Furthermore, the rate capacity and electrolyte permeability of the rGO/NiO films are suggested to be superior, as their capacitances decrease more gradually than those of the NiO films with increasing current density. Electrochemical impedance spectroscopy (EIS) was used to explore this phenomenon by analyzing the conductivity and ion transport behavior of the films. As shown in Figure 6d, the Nyquist profiles of the rGO/NiO and NiO films indicate a lower charge transfer resistance ($R_{ct}$) in rGO/NiO, facilitating faster charge transfer between the electrode and electrolyte, thus enhancing the capacitive performance. [37]

EC smart pixels hold significant promise for information display applications due to their capacity to toggle between various colors and states. This functionality allows displays to dynamically adjust and display content as needed, which is particularly advantageous for complex human–computer interactions such as those found on smartphone screens. Compared to traditional displays, EC smart pixel displays have several benefits: they are low-power displays, thus reducing energy consumption and extending the device battery life; they can be flexible and transparent, opening up new possibilities in display design and application; and they can be produced with low-cost materials and processes, offering a cost-effective alternative for display technology. As EC smart pixel technology evolves, its applications are expected to broaden, and user interfaces are expected to be enhanced, leading to more intuitive and seamless interactions with devices. [22] The $Ti_3C_2T_x$ EC cell demonstrates rapid responsiveness and a significant extinction change per volt (230 nm $V^{-1}$), which is attributed to its low sheet resistance and superior figure of merit (FoMe) compared to those of other MXenes under investigation. The extinction spectral shift, or electro-optical capacitance, is quantified by the slope of the linear change in the plasmon resonance energy with electric potential, as shown in Figure 6e. This figure also illustrates variations in the peak position shift (converted to energy in eV) for different MXene EC systems under various compositions and cathodic potentials. Each MXene composition functions within a distinct wavelength window, and their energy shifts are notably similar: 0.42 eV for $Ti_3C_2T_x$, 0.37 eV for $Ti_2CT_x$, and 0.44 eV for $Ti_{1.6}Nb_{0.4}CT_x$. Additionally, the blueshifts observed during electron injection in electrochemical cycling correspond to changes in the free electron density or carrier concentration ($N_e$), as depicted in Figure 6f. [40]

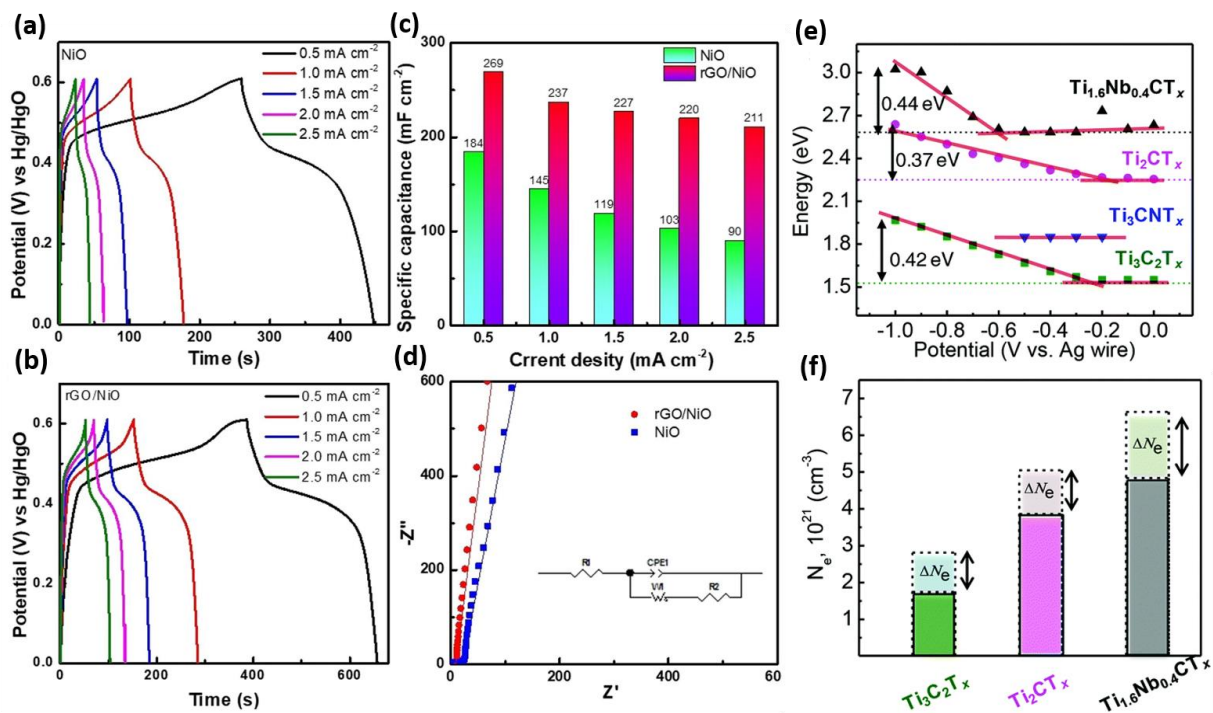

**Figure 6**: ECDs used in different applications. (a, b) GCD curves at different current densities. [37] (c) Calculated specific capacitance at different current densities. [37] (d) Impedance spectra of the equivalent circuit. [37] (e) Variation in the plasmon resonance energy of MXene thin films at cathodic potentials relative to the OCV state (shown by the dotted lines). (f) Estimated relative variation in the electron density of titanium-based MXenes under applied cathodic potentials, i.e., compared between the OCV and −1 V *vs*. Ag wire. [40]

### 4.2 Emerging ECDs in biochemical sensing applications

In recent years, ECDs have also been explored for their potential application in biochemical sensing. EC biochemical sensors are a type of biosensor that use an EC material to detect the presence of specific biomolecules. The color or light absorbance of EC materials changes in response to an applied electric field, which allows these materials to be used as transducers for detecting biomolecular interactions. The sensing mechanism of EC biochemical sensors involves the immobilization of a biomolecule, such as an enzyme or an antibody, on the surface of the EC material. When the biomolecule interacts with its target analyte in the sample, this interaction causes a change in the electric field that results in a change in the color or light absorption of the EC material. The magnitude of the color or absorption change is proportional to the concentration of the target analyte in the sample.

EC biochemical sensors are promising devices with potential for a wide range of applications. The areas of application for ECDs in biochemical sensing include environmental monitoring, food safety, and medical diagnostics. For example, EC sensors could be used to detect the presence of toxic chemicals in water or to monitor the growth of bacteria in food. In medical diagnostics, ECDs can be used to measure the concentrations of various biomarkers, including glucose, lactate, and other biomolecules, in blood or other bodily fluids. One area of research has focused on developing EC sensors for glucose detection. Glucose sensors are critical for the management of diabetes, which is a chronic disease that affects millions of people worldwide. EC sensors can be used to detect glucose by measuring the changes in the color or transparency of the sensors in response to changes in glucose concentration, [41] as shown in Figure 7a. EC biosensors provide a noninvasive and potentially more accurate alternative to traditional diagnostic methods.

EC biochemical sensors have several advantages over traditional biochemical sensors, including a high sensitivity, fast response times, and the ability to operate in real time. However, they also have some limitations. For example, these sensors can be affected by interference from other molecules in the sample, and specialized equipment may be required for measurement and analysis. Additionally, realizing stability and durability of EC materials, particularly in harsh environments, can be challenging. We hope that specific review articles will stimulate greater efforts in studying the basic science of EC sensors and EC materials, which may promote the development of new multifunctional ECDs to fulfill the rising demand for futuristic electronic systems. ECDs have built-in advantages in terms of the signal output, which may be identified based only on the color variation. Combining sensing with EC technologies enables quick visualization of sensory perception data, and the detection results can be directly read based on the observable color of ECDs. This technology has enormous potential for a variety of uses in sensor analysis. [42]

EC biochemical sensors are devices employed to monitor changes in biochemical reactions through EC signals. The first generation of these sensors predominantly relied on oxygen reactions. [43, 44] However, their dependency on the ambient oxygen concentration places significant limitations on their diverse application because variable oxygen levels affect the accuracy and reliability of the sensor. To address the limitations of the first generation of sensors, mediators were incorporated in the second-generation sensors in place of oxygen. These mediators, which maintain steadier levels within the system, facilitate more consistent and reliable measurements across a range of conditions, as shown in Figure 7b. [45] This advancement marked a significant improvement in the adaptability and utility of EC biochemical sensors.

Further enhancements led to the third generation of EC biochemical sensors, in which enzyme denaturation mechanisms are integrated. Often employing iron-based mediators such as ferro/ferricyanide, these sensors demonstrate increased stability and robustness, making them ideal for harsh environments and broader biochemical applications. The third-generation design allows precise monitoring of biological and environmental processes, significantly expanding the practical applications of these sensors. [46, 47] In these advanced sensors, redox enzymes serve as electrocatalysts, enabling direct electron transfer without mediators. [48] This direct electron transfer mechanism necessitates highly effective redox enzymes to facilitate rapid and sensitive biosensor responses (Figure 7c). The electron transfer process in these reactions is crucial for enhancing the reaction efficiency and ensuring charge balance, which is pivotal for maintaining stable and reversible kinetic parameters under constant pH conditions. [49-52]

EC sensors that utilize EC technology are prevalent in biochemical sensing due to their ease of use, affordability, and low power requirements. These sensors deliver colorimetric readings through the detection of color changes in response to biochemical stimuli, enabling the detection of the pH level, various ions such as ammonium, heavy metals, and $Fe(CN)_6^{3/4-}$, and hazardous substances such as chlorpyrifos. An example of this application is the use of a Prussian blue EC electrode for ammonium ion detection, in which color shifts in response to the ion concentration allow accurate sensing (Figure 7d). [53] This illustrates the broad application potential of EC biosensors in ion detection.

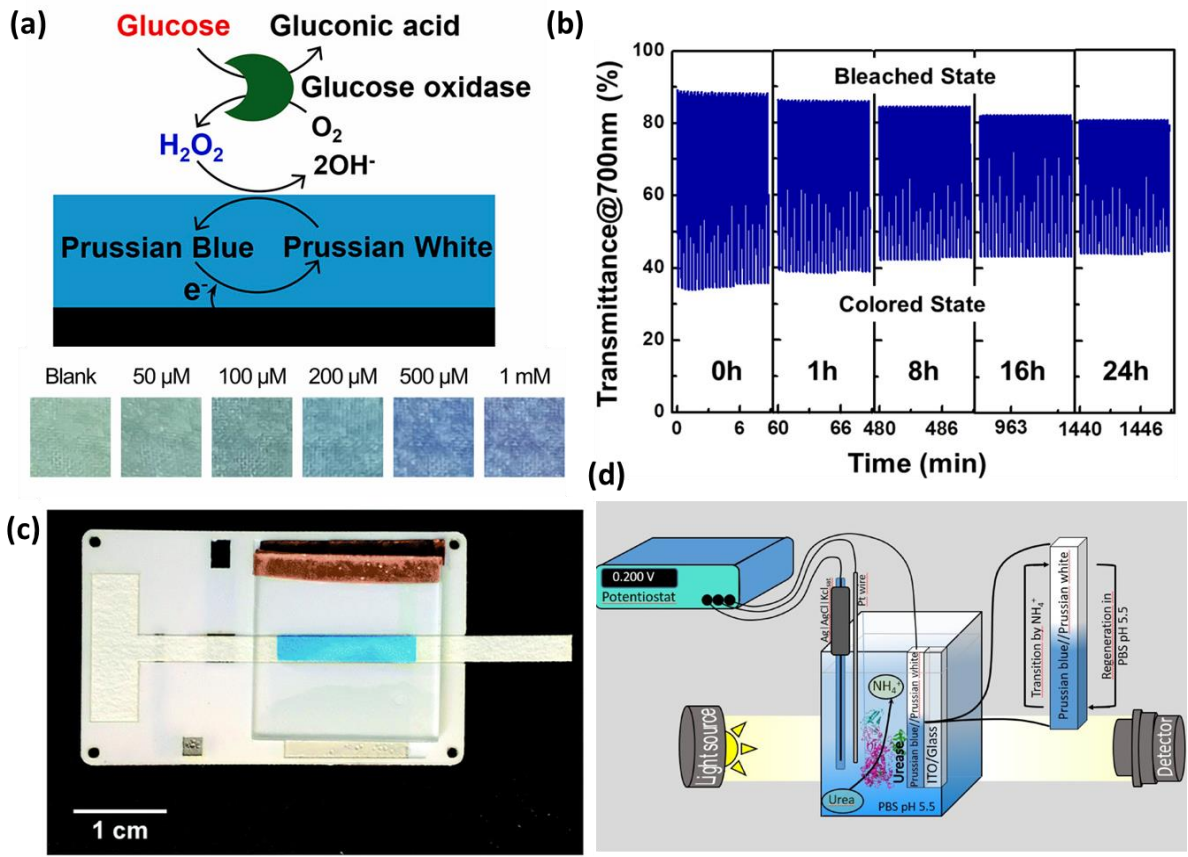

**Figure 7:** ECDs for biochemical sensing. (a) EC glucose sensor and sensing of glucose. [41] (b) Operational stability of the ECD during repeated coloration and bleaching cycles. [45] (c) Photograph of an actual EC biosensor device. [49] (d) Principal scheme of the urea biosensor operation. [53]

### 4.3 ECDs for point-of-care diagnostics

ECDs are poised to significantly impact point-of-care diagnostics due to their intrinsic visual response capabilities and rapid detection times. By integrating EC technology into diagnostic systems, healthcare providers can achieve immediate visual feedback, which is crucial for timely decision-making in clinical settings.

De Matteis et al. introduced an innovative diagnostic device that utilizes the EC properties of $WO_3$ to monitor the sodium (Na) level in human sweat, which serves as a diagnostic indicator for cystic fibrosis (CF). [54] The device operates under a constant applied potential and can switch colors from transparent to various shades of blue through electron injection and Na intercalation into the $WO_3$ structure (Figure 8a). This color change is directly influenced by the concentration of ions intercalated within the EC material. Specifically, a light blue color indicates a Na concentration less than 30 mmol/L, suggesting the absence of CF. In contrast, blue and dark blue intensities correspond to Na levels of 60 mmol/L and 90 mmol/L, respectively, indicating possible and confirmed CF diagnoses (Figure 8b).

Remarkably, this sensor requires only 3 µL of human sweat for analysis, demonstrating its efficiency and potential for point-of-care applications in CF diagnosis, as illustrated in the findings.

Fortunato and colleagues introduced an innovative bioelectrochromic platform for the detection of electrochemically active bacteria (EAB), specifically *Geobacter sulfurreducens* (*Gs*), which is known for its ability to transfer electrons beyond the cell exterior. [55] Utilizing this unique property, the team developed a paper-based EC sensor incorporating $WO_3$ as the EC material (ECM). This sensor leverages the electrochemical activity of *Gs* on paper substrates coated with $WO_3$ nanoparticles through a reducing hydrothermal synthesis method (Figure 8c). The operational principle of the sensor is based on the change in the optical properties of $WO_3$ nanoparticles, which shift between two coloration states—white/yellow and blue—upon the application of an adequate electrochemical potential. The bacteria induce a change in the oxidation state of $WO_3$, visibly turning the sensor blue, thereby indicating positive detection. In this study, the researchers synthesized one-dimensional $WO_3$ nanostructures to maximize the surface area and surface atom density, enhancing interactions with the bacteria. These nanostructures not only improve the sensitivity and specificity of detection but also significantly reduce the response time of EC switching. This reduction in analysis time, combined with an increased contrast ratio, allows precise and readily observable results (Figure 8d). This advancement underscores the potential of integrating bioelectrochromic technology into simple, efficient, and low-cost diagnostic tools for microbial monitoring and detection.

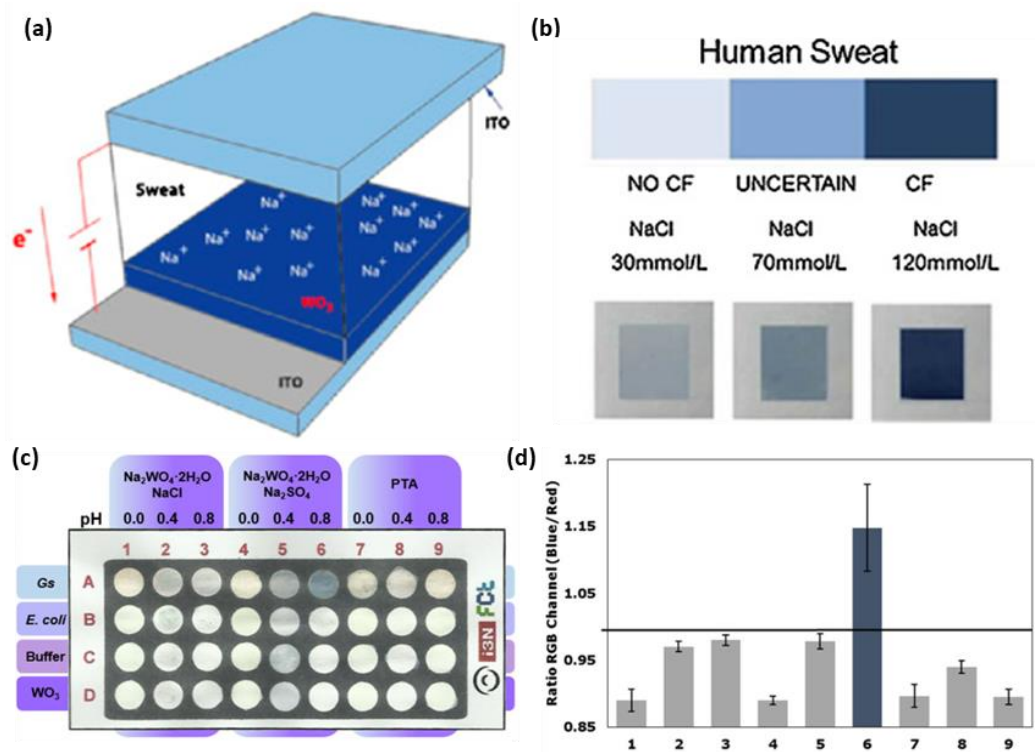

**Figure 8:** ECDs for point-of-care diagnostics. (a) Schematic view of the sweat test device showing the mechanism of sodium intercalation, which induces $WO_3$ coloration. (b) Color variation of EC films depending on the sodium concentration. [54] (c) Photograph of the colorimetric assays of all synthesized WO$_3$ nanoparticles at 5 g/L for the paper-based sensor. (d) RGB analyses for all the samples in contact with *Gs* cells. [55]

## 5. Multifunctional biochemical sensors based on 2D materials

2D materials, such as graphene and TMDs, have emerged as exceptional candidates for biochemical sensing due to their unique physical and chemical properties, such as mechanical robustness, a feasible fluorescence-quenching ability, broadband light absorption, a high carrier mobility, and biocompatibility. These materials are highly sensitive to environmental variations because all their atoms are exposed, enhancing their biochemical sensing capabilities.

Multifunctional biochemical sensors integrate multiple detection capabilities into a single sensor platform, in which 2D materials significantly contribute. These materials can be functionalized with various receptors to simultaneously detect multiple biomarkers, making them ideal for multiplexed assays. With a high surface area and excellent electrical properties, 2D materials sensitively and swiftly detect biomolecules at low concentrations, enabling rapid, real-time monitoring of biochemical processes. These sensors can be miniaturized into portable devices for point-of-care and in-field applications. They support label-free detection by using intrinsic electrical properties, avoiding secondary labels or markers. The unique optical properties, such as the photoluminescence of TMDs, enhance their sensing capabilities. [56] 2D materials are pivotal for detecting disease biomarkers, environmental toxins, and food quality and are essential in biotechnology for monitoring cellular responses and studying protein interactions. Overall, 2D materials enhance the performance and versatility of biosensors, advancing the development of diagnostic and monitoring systems across various sectors.

MoO$_3$ nanostructures are anticipated to exhibit better gas sensing capabilities than bulk micron powder because they have a greater surface-to-volume ratio. Through a specific grinding and sonication procedure, MoO$_3$ nanosheets were exfoliated from bulk α-MoO$_3$ crystals and employed as a gas sensing material to detect alcohol vapor at an ideal temperature of 300 °C. MoS$_2$ nanosheets have recently attracted interest for gas sensing applications because of their simple fabrication and intriguing surface characteristics. According to theoretical research, many groups have suggested that MoSe$_2$ is effective in detecting harmful gases. Due to their unique properties, the sensing capabilities of MoSe$_2$ monolayers for alcohol (methanol and ethanol) were investigated. [57]

The drug acetaminophen (AP), also referred to as paracetamol, is a common analgesic and antipyretic. Detection of AP is crucial because AP overdose can result in severe hepatotoxicity and nephrotoxicity. Since the phenolic hydroxyl group in AP is electrochemically active and can be oxidized, this suggests that AP can be detected electrochemically. The electrocatalytic oxidation of AP on various electrodes modified with two individual materials and their combination indicated that their combination not only integrates their advantages to make the composite film more conductive but also increases its ability to sense the analyte. This investigation was conducted using the cyclic voltammetry (CV) technique. A bare glassy carbon electrode (GCE) and electrodes modified with rGO, poly(3,4-ethylenedioxythiophene) nanotubes (PEDOT NTs), and the rGO-PEDOT NT composite were used to evaluate the electrochemical performance. The initial potential of the rGO-PEDOT NT-modified electrode as a function of pH over the range of 1 to 13 is depicted in the inset of Figure 9a. The slope is calculated to be 59 mV pH$^{-1}$ from the Nernst equation, which is close to the theoretical value of 61 mV pH$^{-1}$ for equal numbers of proton and electron transfer processes. [58]

Chronoamperometry (CA) was used to test the selectivity of $MoS_2/WO_3$/GCE. Figure 9b shows the findings of the selectivity analysis. The obtained results demonstrate that the addition of hydrazine causes a quick increase in the current response but that the performance of $MoS_2/WO_3$/GCE is unaffected by the addition of interfering species. Thus, $MoS_2/WO_3$/GCE is selective for hydrazine sensing. According to electrochemical studies, the $MoS_2/WO_3$ composite has a greater capacity to electrochemically detect hydrazine than virgin $WO_3$ or $MoS_2$. The synergism between $MoS_2$ and $WO_3$ could explain the good rigidity, stability, and reproducibility in its capacity to detect hydrazine. In real sample examinations, $MoS_2/WO_3$/GCE also demonstrated strong recovery, suggesting its potential use in practical applications. [59]

Because graphene oxide (GO) has many desirable characteristics, including high solubility in a wide range of solvents, facile processing, and the presence of oxygen functional groups or defects, it is a good choice for gas sensors. The response of a GO-based sensor can be adjusted by functionalization since the defects or functional groups in GO can serve as reaction sites for gas adsorption, making the gas easily adsorbed on the GO surface and enhancing the selectivity and sensitivity of the sensor. Figure 9c illustrates the electrochemical nitrite sensors created using spindle-shaped $Co_3O_4$ and rGO nanocomposites. These sensors have good stability, high sensitivity, high selectivity, and good repeatability throughout the nitrite detection process. The manufactured sensor demonstrates a linear detection range of 1-380 μM and a detection limit of 0.14 μM.

In addition to the application examples mentioned above, a selection of studies to illustrate the most advanced functionalities of 2D material-based sensors for detecting biochemical analytes is shown in Table 2.

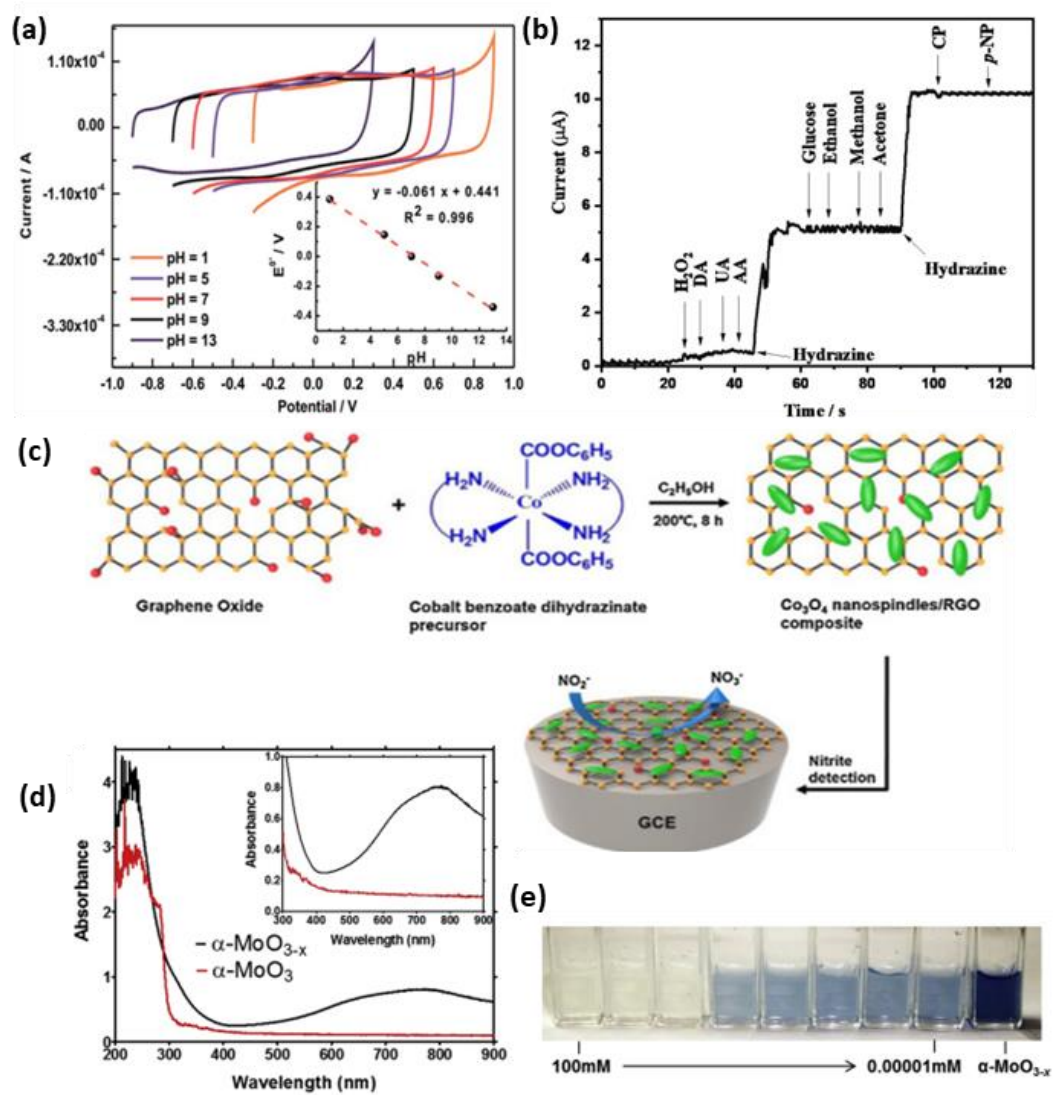

**Figure 9:** Biochemical sensing based on 2D materials. (a) CV curves of the rGO-PEDOT NT-modified electrode evaluated in aqueous buffer solutions at various pH levels; the scan rate was 0.03 V s$^{-1}$. The formal potential plotted against the pH (1–13) is shown in the inset. [58] (b) MoS$_2$, WO$_3$, and GCE with and without interference from other species. Five times more interference-causing chemicals were present than hydrazine. [59] (c) Schematic illustrating the development of an electrochemical nitrite sensor based on Co3O4 and rGO. [60] (d) Reactivity of H$_2$O$_2$ with 2D α-MoO$_{3-x}$ nanoflakes. (e) Change in the color of α-MoO$_{3-x}$ with the H$_2$O$_2$ concentration. [61]

**Table 2:** Selection of studies to illustrate the most advanced functionalities of 2D material-based sensors for detecting biochemical analytes.

| Type of surface interaction and 2D materials | Target biomolecules | Sensing response | | Detection limit | Ref. |
|---|---|---|---|---|---|
| | | Concentration | Relative sensitivity ($\Delta R/R$ or $\Delta I/I$) | | |
| **Tert-butyllithium exfoliated TMDs** | Enzyme | 1–1000 nM | | 2.86 nM | [62] |
| **MoS$_2$ nanosheets and gold nanoparticles** | Catechol C$_6$H$_4$(OH)$_2$ | 2 to 2000 μM | | 2 μM | [51] |
| **MoS$_2$ nanosheet exfoliation** | Lactate oxidase | 0.056 to 0.77 mM | 6.2 μA mM$^{-1}$ | 17 μM | [63] |
| **MoS$_2$-rGO nanocomposite + liquid-assisted exfoliation** | Vi antigen | 0.1 ng mL$^{-1}$ to 1000 ng mL$^{-1}$ | | 100 pg mL$^{-1}$ | [64] |
| **Dielectric-free hydrophobicity of MoS$_2$ + CVD** | Protein & pH | 1 pg/mL | | | [65] |
| **MoS$_2$–polyaniline** | DNA | $10^{-15}$–$10^{-6}$ M | | $2 \times 10^{-16}$ M | [66] |
| **WS$_2$–graphene–Au NPs** | DNA & RNA | 0.01–500 pM | | 0.0023 pM | [67] |
| **WS$_2$–acetylene black** | DNA | 0.001–100 pM | | 0.12 fM | [68] |
| **MoS$_2$–graphene** | Glucose | $10^{-6}$–$10^{-4}$ M | | $2.0 \times 10^{-7}$ M | [69] |

## 6. Toward 2D EC biochemical sensors

EC biosensors, which are known for their visual detection capabilities, are becoming increasingly favored over conventional biosensors. For example, analytical cells respond to biomolecule stimuli, while reporting cells containing ECMs change color in response to electrical signals from the analytical cells. This visual change effectively indicates the presence of biomolecules. [70]

### 6.1 Potential of 2D materials for EC biochemical sensing

2D material-based ECDs have attracted significant interest owing to their unique properties and broad potential applications in technologies such as displays, smart windows, and energy storage systems. The optical properties of ECDs—such as color and transparency—can be altered through the application of an electrical voltage or current. 2D materials are widely used in EC biosensors because they can reversibly shift between blue and nearly transparent, are easy to fabricate, operate at practical voltages, and are biocompatible. The incorporation of 2D materials into ECDs has paved the way for the possibility of biochemical sensing.

Graphene, which is composed of a single layer of carbon atoms, is renowned for its outstanding electrical, mechanical, and optical characteristics. It has been effectively utilized as an electrode material in ECDs, significantly enhancing their performance. Devices that incorporate graphene electrodes exhibit rapid switching speeds, high optical contrast, and robust long-term stability, underscoring the potential of graphene in improving EC technology. [71]

TMDs, such as $MoS_2$ and tungsten diselenide ($WSe_2$), have emerged as excellent candidates for EC applications. These materials display tunable optical properties, which make them suitable for integration into ECDs, either as active layers or electrodes. Compared to devices made with traditional materials, ECDs that employ TMDs have demonstrated superior CE and stability, marking a significant advancement in the field of ECDs. [72]

Black phosphorus (BP), also known as phosphorene, is a notable 2D material in the EC field. BP-based ECDs are distinguished by their broad spectrum of color shifts and robust electrochemical stability. The suitability of BP for flexible substrates makes it ideal for wearable and flexible displays. [73]

In addition, the device performance has been enhanced through the integration of various 2D materials into heterostructures or composites, resulting in synergistic effects. For instance, heterostructures combining graphene with TMDs have been shown to exhibit improved charge transport and EC properties beyond those of the individual materials. [74] Furthermore, composites that include 2D materials with polymers or nanoparticles have been developed to improve their mechanical flexibility and optical qualities. [75]

**6.2 Recent advances in 2D material-based ECDs for biochemical sensing**

The development of ECDs for biochemical sensing has the potential to revolutionize the field of diagnostics by enabling rapid and sensitive detection of a range of biomolecules. Ongoing research in this area is focused on improving the sensitivity and selectivity of EC sensors and on developing practical applications for these devices in clinical and environmental settings.

EC sensors can be designed to detect glucose levels by incorporating enzymes such as glucose oxidase. The enzymatic reaction changes the redox state of the ECM, resulting in a color change proportional to the glucose concentration. pH-sensitive ECMs can be employed for monitoring pH changes in biological samples. The color change indicates the pH variation, making the material useful for application in clinical diagnostics. The immobilization of specific bioreceptors on the EC surface allows selective binding and detection of target analytes. EC sensors offer the advantage of real-time monitoring, enabling continuous and rapid detection of biochemical changes.

As a kind of 2D material, α-MoO3-x nanoflakes can be used as a naked-eye probe for detecting hydrogen peroxide in biological fluids. Figure 9d shows the optical characteristics of α-MoO$_{3-x}$, H$_2$O$_2$, and their mixture. A broad peak between 600 and 900 nm is visible in the optical spectrum of α-MoO$_{3-x}$. The characteristic deep blue color of the solution is caused by the typical absorption of α-MoO$_{3-x}$ in the visible region (600-1000 nm). Interestingly, in the presence of H$_2$O$_2$, the intensity of the peak in the visible region decreases, while the intensity of the peak in the UV region increases. This distinctive result indicates that the oxidized form of α-MoO$_3$ is present. Figure 9e displays an image of the color variations caused by H$_2$O$_2$ taken under ambient light. This figure demonstrates how quickly and effectively H$_2$O$_2$ can be detected with the naked eye. [61]

## 6.3 Challenges and outlook

The ongoing development of ECDs is driven by the demand for energy-efficient, dynamically adjustable optical devices suitable for biochemical sensors. Continued advancements in the study of ECMs and device architectures are expected to yield ECDs with improved performance and expanded functionalities.

The atomic composition of 2D materials can be precisely engineered and doped, introducing a level of customization that allows sensor properties to be tailored to target specific biochemical markers or analytes with high affinity and selectivity. This capability not only enhances the performance of ECD-based sensors but also expands their applicability across various domains of health monitoring, environmental sensing, and even industrial process control.

The versatility of 2D materials, combined with the visual detection capabilities of ECDs, enables the development of noninvasive, real-time, and highly efficient sensing platforms. These sensors can be designed to detect a wide range of biological and chemical substances, from ions and small molecules to larger biomolecules such as proteins and enzymes, making them ideal for applications in personalized medicine, wearable health monitoring, and point-of-care diagnostics.

However, several ongoing problems, such as morphological control, quality/quantity regulation, and preparation cost issues, continue to present significant difficulties for researchers. For the production of high-performance ECDs, the relevant deposition technology and/or film-forming technology must be improved in addition to the nanomaterial preparation technology. For different kinds of transdisciplinary ECDs, integration methods, design guidelines, and performance improvements are highlighted.

One of the primary challenges in developing ECDs is achieving a high CE, which is defined as the ratio of the change in the OD to the charge injected into the device. A high CE is crucial for ensuring high contrast and minimizing the energy required for device operation. To enhance this efficiency, various strategies have been implemented, such as utilizing nanostructured ECMs, incorporating ion-conducting polymers, and optimizing the device architecture.

Another significant challenge is ensuring the long-term stability and durability of ECDs. ECMs are prone to degradation over time due to the cyclic nature of the electric field, exposure to environmental factors such as moisture and heat, and chemical interactions within the device. To combat these issues, research has focused on developing materials with enhanced stability and employing encapsulation techniques to shield ECLs from external elements.

Significant research efforts have been made in novel device design, comprehensive electrochemical kinetic analysis, and advanced synthetic methods using 2D materials for transparent conductors and EC films. These elements are considered pivotal for significantly enhancing the EC performance and expanding the applicability of ECDs. For the development of high-performance, versatile ECDs, synthetic techniques are employed to fabricate porous films that provide near-perfect optical modulation and rapid switching capabilities. Various methods have been utilized to produce high-performance EC films, including hydrothermal/inkjet printing, sol-gel/inkjet printing, integration of hybrid transparent conductors with ECLs, and electron beam evaporation. These innovations highlight the ongoing advancements in the field and underscore the potential for broader application and higher efficiency of ECDs.

## 7. Summary and Concluding Remarks

In this review, we comprehensively explored sensing technologies incorporating EC technology, highlighting the significant progress and rapid developments in the 2D EC sector in recent years.

EC sensors have revolutionized the detection process by enabling direct observation of the results through color changes in ECDs. The integration of 2D materials into ECD biochemical sensors represents a promising frontier in sensor technology. The visual aspect of ECDs, combined with their ability to be engineered for specificity and sensitivity using advanced materials such as 2D nanostructures, enhances their applicability for detecting a wide array of biochemical markers directly at the patient's side. Furthermore, ECD-based sensors are highly versatile and can be designed to be compact, portable, and easy to use, making them ideal for point-of-care applications when complex laboratory facilities are unavailable. Their low power requirements and potential for integration into flexible and wearable formats further underscore their suitability for modern healthcare environments, in which rapid, accurate, and user-friendly diagnostic tools are increasingly vital.

In summary, in this review, we comprehensively summarized sensing technologies with 2D EC technology. The incorporation of 2D materials into ECD-based biochemical sensors has the potential to revolutionize the fields of diagnostics and sensing, driven by their atomic-level customization, large surface-to-volume ratios, and inherent visual feedback mechanism. Incredible advancements have been made in the 2D EC area, which has experienced rapid growth in recent years. This review also covered recent advancements in sensor technology utilizing 2D materials, from morphological aspects to device engineering. The compositional and structural attributes of 2D nanostructures, along with their surface properties, fundamentally influence their electrical characteristics, opening up exciting possibilities for EC sensor applications.

**Conflicts of Interest:** The authors declare no conflicts of interest.

**Acknowledgments:** The research leading to this article has received much appreciated funding from the National Natural Science Foundation of China (No. 52073160), the National Key Research and Development Program of China (No. 2020YFF01014706), and the Beijing Municipal Science and Technology Commission (Z211100002421012).

**Data Availability:** All the data generated or analyzed during this study have been included in this published article.

## References

1. Cai, Y.; Yang, B.; Ji, J.; Sun, F.; Zhao, Y.; Yu, L.; Zhao, C.; Liu, M.; Liu, M.; He, Y. J. A. M. T., A Universal Tandem Device of DC‐Driven Electrochromism and AC‐Driven Electroluminescence for Multi‐Functional Smart Windows. *Advanced Materials Technologies* **2022**, 2201682.
2. Granqvist, C. G. J. T. s. f., Electrochromics for smart windows: Oxide-based thin films and devices. *Thin Solid Films* **2014,** 564, 1-38.
3. Balendhran, S.; Walia, S.; Nili, H.; Ou, J. Z.; Zhuiykov, S.; Kaner, R. B.; Sriram, S.; Bhaskaran, M.; Kalantar‐zadeh, K., Two‐dimensional molybdenum trioxide and dichalcogenides. *Advanced Functional Materials* **2013,** 23, (32), 3952-3970.
4. Sharma, R.; Sharma, M.; Goswamy, J. J. I. J. o. E. R., Synthesis and characterization of MoS2/WO3 nanocomposite for electrochromic device application. *International Journal of Energy Research* **2022**.
5. Chen, X.; Zhang, H.; Li, W.; Xiao, Y.; Ge, Z.; Li, Y.; Zhang, X. J. M. L., Electro-optical performance of all solid state electrochromic devices with NaF electrolytes. *Materials Letters* **2022,** 324, 132692.

6. Zohrevand, N.; Madrakian, T.; Ghoorchian, A.; Afkhami, A. J. E. A., Simple electrochromic sensor for the determination of amines based on the proton sensitivity of polyaniline film. *Electrochimica Acta* **2022,** 427, 140856.
7. Bi, S.; Jin, W.; Han, X.; Cao, X.; He, Z.; Asare-Yeboah, K.; Jiang, C. J. N. E., Ultra-fast-responsivity with sharp contrast integrated flexible piezo electrochromic based tactile sensing display. *Nano Energy* **2022,** 102, 107629.
8. Xiao, M.; Wei, S.; Chen, J.; Tian, J.; Brooks Iii, C. L.; Marsh, E. N. G.; Chen, Z., Molecular Mechanisms of Interactions between Monolayered Transition Metal Dichalcogenides and Biological Molecules. *Journal of the American Chemical Society* **2019,** 141, (25), 9980-9988.
9. Rao, C.; Gopalakrishnan, K.; Maitra, U. J. A. a. m.; interfaces, Comparative study of potential applications of graphene, $MoS_2$, and other two-dimensional materials in energy devices, sensors, and related areas. *ACS applied materials & interfaces* **2015,** 7, (15), 7809-7832.
10. Li, Y.; Yang, B.; Xu, S.; Huang, B.; Duan, W. J. A. A. E. M., Emergent phenomena in magnetic two-dimensional materials and van der waals heterostructures. *ACS Applied Electronic Materials* **2022,** 4, (7), 3278-3302.
11. Mphuthi, N.; Sikhwivhilu, L.; Ray, S. S. J. B., Functionalization of 2D $MoS_2$ Nanosheets with Various Metal and Metal Oxide Nanostructures: Their Properties and Application in Electrochemical Sensors. *Biosensors* **2022,** 12, (6), 386.
12. Li, T.; Shang, D.; Gao, S.; Wang, B.; Kong, H.; Yang, G.; Shu, W.; Xu, P.; Wei, G. J. B., Two-dimensional material-based electrochemical sensors/biosensors for food safety and biomolecular detection. *Biosensors* **2022,** 12, (5), 314.
13. Kalia, S.; Rana, D. S.; Thakur, N.; Singh, D.; Kumar, R.; Singh, R. K. J. M. C.; Physics, Two-dimensional layered molybdenum disulphide ($MoS_2$)-reduced graphene oxide (rGO) heterostructures modified with $Fe_3O_4$ for electrochemical sensing of epinephrine. *Materials Chemistry and Physics* **2022**, 126274.
14. Jiao, L.; Xu, W.; Wu, Y.; Yan, H.; Gu, W.; Du, D.; Lin, Y.; Zhu, C. J. C. S. R., Single-atom catalysts boost signal amplification for biosensing. *Chemical Society Reviews* **2021,** 50, (2), 750-765.
15. Zribi, R.; Foti, A.; Donato, M. G.; Gucciardi, P. G.; Neri, G. J. E. A., Electrochemical and sensing properties of 2D-$MoS_2$ nanosheets produced via liquid cascade centrifugation. *Electrochimica Acta* **2022,** 436, 141433.
16. Iqbal, M. A.; Malik, M.; Shahid, W.; Ahmad, W.; Min-Dianey, K. A.; Pham, P. V. J. s. C. N. M. P., Chemistry, Classification,; Emerging Applications in Industry, B.; Agriculture, Plasmonic 2D Materials: Overview, Advancements, Future Prospects and Functional Applications. In *21st Century Nanostructured Materials - Physics, Chemistry, Classification, and Emerging Applications in Industry, Biomedicine, and Agriculture*, IntechOpen: 2022; p 47.
17. Choi, W.; Choudhary, N.; Han, G. H.; Park, J.; Akinwande, D.; Lee, Y. H., Recent development of two-dimensional transition metal dichalcogenides and their applications. *Materials Today* **2017,** 20, (3), 116-130.
18. Wang, Z.; Zhu, W.; Qiu, Y.; Yi, X.; von dem Bussche, A.; Kane, A.; Gao, H.; Koski, K.; Hurt, R., Biological and environmental interactions of emerging two-dimensional nanomaterials. *Chemical Society Reviews* **2016,** 45, (6), 1750-80.
19. Singh, N. B.; Hua Su, C.; Arnold, B.; Choa, F.-S.; Sova, S.; Cooper, C., Multifunctional 2D-Materials: Gallium Selenide. *Materials Today: Proceedings* **2017,** 4, (4), 5471-5477.
20. Zhou, J.; Shen, L.; Costa, M. D.; Persson, K. A.; Ong, S. P.; Huck, P.; Lu, Y.; Ma, X.; Chen, Y.; Tang, H. J. S. d., 2DMatPedia, an open computational database of two-dimensional materials from top-down and bottom-up approaches. *Scientific data* **2019,** 6, (1), 86.
21. Li, M.; Wu, Z.; Tian, Y.; Pan, F.; Gould, T.; Zhang, S. J. A. M. T., Nanoarchitectonics of Two‐Dimensional Electrochromic Materials: Achievements and Future Challenges. *Advanced Materials Interfaces* **2022**, 2200917.
22. Gu, C.; Jia, A.-B.; Zhang, Y.-M.; Zhang, S. X.-A. J. C. R., Emerging electrochromic materials and devices for future displays. *Chemical Reviews* **2022,** 122, (18), 14679-14721.

23. Tang, X.; Chen, G.; Li, Z.; Li, H.; Zhang, Z.; Zhang, Q.; Ou, Z.; Li, Y.; Qi, C.; Luo, J., Structure evolution of electrochromic devices from 'face-to-face' to 'shoulder-by-shoulder'. *Journal of Materials Chemistry C* **2020,** 8, (32), 11042-11051.
24. Pellitero, M. A.; del Campo, F. J., Electrochromic sensors: Innovative devices enabled by spectroelectrochemical methods. *Current Opinion in Electrochemistry* **2019,** 15, 66-72.
25. Kandpal, S.; Ghosh, T.; Rani, C.; Rani, S.; Pathak, D. K.; Tanwar, M.; Bhatia, R.; Sameera, I.; Kumar, R. J. S. E. M.; Cells, S., MoS2 nano-flower incorporation for improving organic-organic solid state electrochromic device performance. *Solar Energy Materials and Solar Cells* **2022,** 236, 111502.
26. Chen, W. H.; Li, F. W.; Liou, G. S., Novel Stretchable Ambipolar Electrochromic Devices Based on Highly Transparent AgNW/PDMS Hybrid Electrodes. *Advanced Optical Materials* **2019,** 7, (19).
27. Polat, E. O.; Balcı, O.; Kocabas, C. J. S. R., Graphene based flexible electrochromic devices. *Scientific Reports* **2014,** 4, (1), 6484.
28. Wang, Y.; Niu, H.; Lu, Q.; Zhang, W.; Qiao, X.; Niu, H.; Zhang, Y.; Wang, W. J. S. A. P. A. M.; Spectroscopy, B., From aerospace to screen: Multifunctional poly (benzoxazine) s based on different triarylamines for electrochromic, explosive detection and resistance memory devices. *Spectrochimica Acta Part A: Molecular and Biomolecular Spectroscopy* **2020,** 225, 117524.
29. Chen, F.; Fu, X.; Zhang, J.; Wan, X. J. E. A., Near-infrared and multicolored electrochromism of solution processable triphenylamine-anthraquinone imide hybrid systems. *Electrochimica Acta* **2013,** 99, 211-218.
30. Rai, V.; Singh, R. S.; Blackwood, D. J.; Zhili, D. J. A. E. M., A review on recent advances in electrochromic devices: a material approach. *Advanced Engineering Materials* **2020,** 22, (8), 2000082.
31. Rakibuddin, M.; Kim, H., Fabrication of MoS2/WO3 nanocomposite films for enhanced electro-chromic performance. *New Journal of Chemistry* **2017,** 41, (24), 15327-15333.
32. Yu, S.; Wu, X.; Wang, Y.; Guo, X.; Tong, L. J. A. M., 2D materials for optical modulation: challenges and opportunities. *Advanced Materials* **2017,** 29, (14), 1606128.
33. Ahmad, K.; Shinde, M. A.; Song, G.; Kim, H., Design and fabrication of MoSe2/WO3 thin films for the construction of electrochromic devices on indium tin oxide based glass and flexible substrates. *Ceramics International* **2021,** 47, (24), 34297-34306.
34. Gadgil, B.; Damlin, P.; Heinonen, M.; Kvarnström, C., A facile one step electrostatically driven electrocodeposition of polyviologen–reduced graphene oxide nanocomposite films for enhanced electrochromic performance. *Carbon* **2015,** 89, 53-62.
35. Novak, T. G.; Kim, J.; Tiwari, A. P.; Kim, J.; Lee, S.; Lee, J.; Jeon, S., 2D MoO3 Nanosheets Synthesized by Exfoliation and Oxidation of MoS2 for High Contrast and Fast Response Time Electrochromic Devices. *ACS Sustainable Chemistry & Engineering* **2020,** 8, (30), 11276-11282.
36. Rakibuddin, M.; Shinde, M. A.; Kim, H. J. E. A., Facile sol–gel fabrication of MoS2 bulk, flake and quantum dot for electrochromic device and their enhanced performance with WO3. *Electrochimica Acta* **2020,** 349, 136403.
37. Xue, J.; Xu, H.; Wang, S.; Hao, T.; Yang, Y.; Zhang, X.; Song, Y.; Li, Y.; Zhao, J. J. A. S. S., Design and synthesis of 2D rGO/NiO heterostructure composites for high-performance electrochromic energy storage. *Applied Surface Science* **2021,** 565, 150512.
38. Zhao, S.; Huang, W.; Guan, Z.; Jin, B.; Xiao, D. J. E. A., A novel bis (dihydroxypropyl) viologen-based all-in-one electrochromic device with high cycling stability and coloration efficiency. *Electrochimica Acta* **2019,** 298, 533-540.
39. Eh, A. L. S.; Tan, A. W. M.; Cheng, X.; Magdassi, S.; Lee, P. S. J. E. T., Recent advances in flexible electrochromic devices: prerequisites, challenges, and prospects. *Energy Technology* **2018,** 6, (1), 33-45.

40. Valurouthu, G.; Maleski, K.; Kurra, N.; Han, M.; Hantanasirisakul, K.; Sarycheva, A.; Gogotsi, Y. J. N., Tunable electrochromic behavior of titanium-based MXenes. *Nanoscale* **2020**, 12, (26), 14204-14212.
41. Yeon, S. Y.; Seo, M.; Kim, Y.; Hong, H.; Chung, T. D., Paper-based electrochromic glucose sensor with polyaniline on indium tin oxide nanoparticle layer as the optical readout. *Biosensors and Bioelectronics* **2022**, 203, 114002.
42. Yang, P.; Sun, P.; Mai, W., Electrochromic energy storage devices. *Materials Today* **2016**, 19, (7), 394-402.
43. Xu, L.; Li, D.; Ramadan, S.; Li, Y.; Klein, N., Facile biosensors for rapid detection of COVID-19. *Biosensors and Bioelectronics* **2020**, 170, 112673.
44. Porcel-Valenzuela, M.; Ballesta-Claver, J.; de Orbe-Payá, I.; Montilla, F.; Capitán-Vallvey, L. F., Disposable electrochromic polyaniline sensor based on a redox response using a conventional camera: A first approach to handheld analysis. *Journal of Electroanalytical Chemistry* **2015**, 738, 162-169.
45. Yun, T. Y.; Li, X.; Bae, J.; Kim, S. H.; Moon, H. C., Non-volatile, Li-doped ion gel electrolytes for flexible WO3-based electrochromic devices. *Materials & Design* **2019**, 162, 45-51.
46. Eggins, B. R., *Chemical sensors and biosensors*. John Wiley & Sons: 2002; Vol. 2.
47. Palenzuela, J.; Vinuales, A.; Odriozola, I.; Cabanero, G.; Grande, H. J.; Ruiz, V., Flexible viologen electrochromic devices with low operational voltages using reduced graphene oxide electrodes. *ACS Applied materials & interfaces* **2014**, 6, (16), 14562-7.
48. Kuznetsov, B.; Shumakovich, G.; Koroleva, O.; Yaropolov, A., On applicability of laccase as label in the mediated and mediatorless electroimmunoassay: effect of distance on the direct electron transfer between laccase and electrode. *Biosensors and Bioelectronics* **2001**, 16, (1-2), 73-84.
49. Pellitero, M. A.; Guimera, A.; Kitsara, M.; Villa, R.; Rubio, C.; Lakard, B.; Doche, M. L.; Hihn, J. Y.; Javier Del Campo, F., Quantitative self-powered electrochromic biosensors. *Chemical science* **2017**, 8, (3), 1995-2002.
50. Ghindilis, A. L.; Atanasov, P.; Wilkins, E., Enzyme‐catalyzed direct electron transfer: Fundamentals and analytical applications. *Electroanalysis* **1997**, 9, (9), 661-674.
51. Zhang, Y.; Li, X.; Li, D.; Wei, Q., A laccase based biosensor on AuNPs-MoS2 modified glassy carbon electrode for catechol detection. *Colloids and Surfaces B: Biointerfaces* **2020**, 186, 110683.
52. Fang, A.; Ng, H. T.; Li, S. F. Y., A high-performance glucose biosensor based on monomolecular layer of glucose oxidase covalently immobilised on indium–tin oxide surface. *Biosensors and Bioelectronics* **2003**, 19, (1), 43-49.
53. Valiūnienė, A.; Virbickas, P.; Medvikytė, G.; Ramanavičius, A. J. E., Urea biosensor based on electrochromic properties of Prussian blue. *Electroanalysis* **2020**, 32, (3), 503-509.
54. De Matteis, V.; Cannavale, A.; Blasi, L.; Quarta, A.; Gigli, G., Chromogenic device for cystic fibrosis precocious diagnosis: A "point of care" tool for sweat test. *Sensors and Actuators B: Chemical* **2016**, 225, 474-480.
55. Marques, A. C.; Santos, L.; Costa, M. N.; Dantas, J. M.; Duarte, P.; Gonçalves, A.; Martins, R.; Salgueiro, C. A.; Fortunato, E., Office Paper Platform for Bioelectrochromic Detection of Electrochemically Active Bacteria using Tungsten Trioxide Nanoprobes. *Scientific Reports* **2015**, 5, (1), 9910.
56. Wang, S.; Liu, Y.; Zhu, A.; Tian, Y. J. A. C., In vivo electrochemical biosensors: Recent advances in molecular design, electrode materials, and electrochemical devices. *Analytical Chemistry* **2023**, 95, (1), 388-406.
57. Jha, R. K.; Bhat, N. J. A. M. I., Recent progress in chemiresistive gas sensing technology based on molybdenum and tungsten chalcogenide nanostructures. *Advanced Materials Interfaces* **2020**, 7, (7), 1901992.

58. Huang, T.-Y.; Kung, C.-W.; Wei, H.-Y.; Boopathi, K. M.; Chu, C.-W.; Ho, K.-C., A high performance electrochemical sensor for acetaminophen based on a rGO–PEDOT nanotube composite modified electrode. *Journal of Materials Chemistry A* **2014,** 2, (20), 7229-7237.
59. Ahmad, K.; Kim, H. J. M. S. i. S. P., Synthesis of MoS2/WO3 hybrid composite for hydrazine sensing applications. *Materials Science in Semiconductor Processing* **2022,** 148, 106803.
60. Haldorai, Y.; Kim, J. Y.; Vilian, A. E.; Heo, N. S.; Huh, Y. S.; Han, Y.-K. J. S.; Chemical, A. B., An enzyme-free electrochemical sensor based on reduced graphene oxide/Co3O4 nanospindle composite for sensitive detection of nitrite. *Sensors and Actuators B: Chemical* **2016,** 227, 92-99.
61. Sharma, A. K.; Pandey, S.; Sharma, K. H.; Nerthigan, Y.; Khan, M. S.; Hang, D.-R.; Wu, H.-F. J. A. c. a., Two dimensional α-MoO3-x nanoflakes as bare eye probe for hydrogen peroxide in biological fluids. *Analytica chimica acta* **2018,** 1015, 58-65.
62. Nasir, M. Z. M.; Mayorga-Martinez, C. C.; Sofer, Z. k.; Pumera, M., Two-dimensional 1T-phase transition metal dichalcogenides as nanocarriers to enhance and stabilize enzyme activity for electrochemical pesticide detection. *ACS nano* **2017,** 11, (6), 5774-5784.
63. Parra-Alfambra, A. M.; Casero, E.; Vázquez, L.; Quintana, C.; del Pozo, M.; Petit-Domínguez, M. D., MoS2 nanosheets for improving analytical performance of lactate biosensors. *Sensors and Actuators B: Chemical* **2018,** 274, 310-317.
64. Pathania, P. K.; Saini, J. K.; Vij, S.; Tewari, R.; Sabherwal, P.; Rishi, P.; Suri, C. R., Aptamer functionalized MoS2-rGO nanocomposite based biosensor for the detection of Vi antigen. *Biosensors and Bioelectronics* **2018,** 122, 121-126.
65. Lee, J.; Dak, P.; Lee, Y.; Park, H.; Choi, W.; Alam, M. A.; Kim, S., Two-dimensional layered MoS 2 biosensors enable highly sensitive detection of biomolecules. *Scientific reports* **2014,** 4, (1), 1-7.
66. Li, J.; Yan, L.; Tang, X.; Feng, H.; Hu, D.; Zha, F., Robust superhydrophobic fabric bag filled with polyurethane sponges used for vacuum‐assisted continuous and ultrafast absorption and collection of oils from water. *Advanced Materials Interfaces* **2016,** 3, (9), 1500770.
67. Huang, K.-J.; Liu, Y.-J.; Wang, H.-B.; Gan, T.; Liu, Y.-M.; Wang, L.-L., Signal amplification for electrochemical DNA biosensor based on two-dimensional graphene analogue tungsten sulfide–graphene composites and gold nanoparticles. *Sensors Actuators B: Chemical* **2014,** 191, 828-836.
68. Shuai, H.-L.; Huang, K.-J.; Chen, Y.-X., A layered tungsten disulfide/acetylene black composite based DNA biosensing platform coupled with hybridization chain reaction for signal amplification. *Journal of Materials Chemistry B* **2016,** 4, (6), 1186-1196.
69. Hu, Y.; Huang, Y.; Tan, C.; Zhang, X.; Lu, Q.; Sindoro, M.; Huang, X.; Huang, W.; Wang, L.; Zhang, H., Two-dimensional transition metal dichalcogenide nanomaterials for biosensing applications. *Materials Chemistry Frontiers* **2017,** 1, (1), 24-36.
70. Fan, H.; Wei, W.; Hou, C.; Zhang, Q.; Li, Y.; Li, K.; Wang, H., Wearable electrochromic materials and devices: from visible to infrared modulation. *Journal of Materials Chemistry C* **2023,** 11, (22), 7183-7210.
71. Köç Bakacak, P.; Kovalska, E.; Tüzemen, S., Graphene for switchable flexible smart windows application. *Optical Materials* **2024,** 151, 115302.
72. Ma, D.; Wang, J.; Wei, H.; Guo, Z., Electrochromic Materials and Devices: Fundamentals and Nanostructuring Approaches. In *Multifunctional Nanocomposites for Energy and Environmental Applications*, 2018; pp 231-270.
73. Xiong, J.; Cui, P.; Chen, X.; Wang, J.; Parida, K.; Lin, M.-F.; Lee, P. S., Skin-touch-actuated textile-based triboelectric nanogenerator with black phosphorus for durable biomechanical energy harvesting. *Nature Communications* **2018,** 9, (1), 4280.
74. Maggini, L.; Ferreira, R. R., 2D material hybrid heterostructures: achievements and challenges towards high throughput fabrication. *Journal of Materials Chemistry C* **2021,** 9, (44), 15721-15734.

75. Shanmugam, V.; Mensah, R. A.; Babu, K.; Gawusu, S.; Chanda, A.; Tu, Y.; Neisiany, R. E.; Försth, M.; Sas, G.; Das, O., A Review of the Synthesis, Properties, and Applications of 2D Materials. *Particle & Particle Systems Characterization* **2022,** 39, (6), 2200031.